\documentclass[aps,prc,showpacs,twocolumn,superscriptaddress,nofootinbib,preprintnumbers]{revtex4-1}

\usepackage{longtable}
\usepackage{graphicx}
\usepackage{dcolumn}
\usepackage{bm}
\usepackage{amsmath,amssymb}

\newcommand{\re}{\text{Re }}
\newcommand{\im}{\text{Im }}

\usepackage[normalem]{ulem}  
\usepackage[dvips]{color} 

\renewcommand\sout{\bgroup \color{red} \ULdepth=-.5ex \ULset}

\begin{document}

\preprint{YITP-17-85}


\title{Universal physics of the few-body system of two neutrons and one flavored meson}


\author{Udit Raha}
\email[]{udit.raha@iitg.ac.in}
\affiliation{Department of Physics, Indian Institute of Technology Guwahati, 781 039 Assam, India}
\author{Yuki~Kamiya}
\email[]{yuki.kamiya@yukawa.kyoto-u.ac.jp}
\affiliation{Yukawa Institute for Theoretical Physics, Kyoto University, Kyoto 606-8502, Japan}
\author{Shung-Ichi Ando}
\email[]{sando@sunmoon.ac.kr}
\affiliation{Department of Information Communication and Display Engineering, 
Sunmoon University, Asan, Chungnam 31460, Korea}
\author{Tetsuo~Hyodo}
\email[]{hyodo@yukawa.kyoto-u.ac.jp}
\affiliation{Yukawa Institute for Theoretical Physics, Kyoto University, Kyoto 606-8502, Japan}


\date{\today}

\begin{abstract}
We investigate the $s$-wave three-body system of two neutrons and one flavored meson with total spin-isospin 
$J=0,I=3/2$. The meson-neutron scattering length can become infinitely large when extrapolated to an unphysical region 
of the quark mass between strangeness and charm in the so-called zero coupling limit. Using a low-energy cluster effective 
field theory, we demonstrate that the Efimov effect is formally manifest in the three-body system when the meson-neutron 
scattering length approaches the unitary limit of the two-body interaction. We thereby discuss the consequence of remnant 
universal physics in the physical $K^{-}nn$ and $D^{0}nn$ systems. 
\end{abstract}




\maketitle

\section{Introduction}

Few-body systems of an antikaon and nucleons are of great interest in the strangeness nuclear 
physics~\cite{Hyodo:2011ur,Gal:2016boi}. The existence of the $\Lambda(1405)$ resonance below 
the $\bar{K}N$ threshold implies that the $\bar{K}N$ interaction in the isospin $I=0$ channel 
is strongly attractive with which the antikaon could be bound in nuclei~\cite{PL7.288,Akaishi:2002bg}. 
Recently, three-body $\bar{K}NN$ systems have been a matter of intensive investigation based on rigorous 
few-body techniques~\cite{Shevchenko:2006xy,Shevchenko:2007ke,Ikeda:2007nz,Ikeda:2008ub,Yamazaki:2007cs,Dote:2008in,Dote:2008hw,Wycech:2008wf,Ikeda:2010tk,Barnea:2012qa,Ohnishi:2017uni,Revai:2016muw,Hoshino:2017mty}. 
To maximize the $I=0$ contribution of the $\bar{K}N$ pair in the $s$-wave $\bar{K}NN$ system, the state 
with the total spin $J=0$ and isospin $I=1/2$ is mainly studied. It is agreed among several different groups 
that the $J=0$ system supports a quasi-bound state below the threshold, although the quantitative predictions 
are not yet well converged. There are some studies of the state with $J=1$ and $I=1/2$, but the quasi-bound 
state is not found below the $\Lambda^{*}N$ threshold in Refs.~\cite{Uchino:2011jt,Barnea:2012qa}, presumably 
because of the small fraction of the $I=0$ component of the $\bar{K}N$ pair. In Ref.~\cite{Wycech:2008wf}, it 
is shown that the $J=2$ and $I=3/2$ state is bound because of the $p$-wave $\bar{K}N$ interaction that generates 
the $\Sigma(1385)$ state.

Currently, nuclei with a heavy flavor (charm and bottom) meson is a subject of increasing 
attention~\cite{Hosaka:2016ypm}. This is triggered by the observation of many new heavy quarkonium-like 
hadronic states, so-called the $XYZ$ states, above the open-charm or open-bottom 
thresholds~\cite{Brambilla:2010cs,Hosaka:2016pey,Olive:2016xmw}. Because the near-threshold states are 
often interpreted as {\it hadronic molecular states}, the existence of the $XYZ$ states indicates that 
the heavy flavor meson could be a constituent in the formation of such exotic hadronic bound states. 
To assess the possible existence of the bound heavy mesons in nuclei, two-body $DN$ 
interactions~\cite{Hofmann:2005sw,Mizutani:2006vq,GarciaRecio:2008dp,Haidenbauer:2010ch,Yamaguchi:2011xb,Yamaguchi:2013ty} 
and three-body quasi-bound systems~\cite{Bayar:2012dd,Yamaguchi:2013hsa} involving $D$ mesons have been 
studied extensively. With a plethora of high-precision data currently available from a host 
of modern experimental facilities, such as BaBar, Belle, CDF, D0, BES, CMS, and LHCb, similar investigations 
are also being extended to the open-bottom sector.

In general, the dynamics of a three-body problem reflect subtle details of two-body interactions. 
In some circumstances, gross properties of three-body systems can be assessed in terms of a few parameters 
which solely characterize the nature of two-body interactions. In fact, when the two-body scattering 
length $a$ is much larger than the interaction range $r_0$, microscopic details of two-body interactions 
become irrelevant as a consequence of low-energy {\it universality}~\cite{Braaten:2004rn,Naidon:2016dpf}. 
Universal physics can manifest themselves in sharply contrasting manner in the two- and three-body sectors. 
In the two-body sector, universal predictions are quite simple, manifested, e.g., as a shallow two-body bound 
state {\it dimer} for $a>0$ with the eigenenergy in the scaling limit given by $E_{\rm dimer}=-1/(2\mu_{\rm red} a^2)$, 
where $\mu_{\rm red}$ is the generic two-body reduced mass. However, universal predictions can become more complex 
in the three-body sector. The most striking phenomenon in a quantum mechanical three-body problem is the so-called 
{\it Efimov effect}~\cite{Efimov:1970zz}, characterized by the emergence of an infinitely many and arbitrary 
shallow geometrically spaced bound states with accumulation point at zero energy in $s$-wave three-body systems. 
In the context of hadron physics, the Efimov effect and low-energy universality have been discussed in  
three-nucleon systems~\cite{Braaten:2003eu,Konig:2016utl,Canham:2009zq}, charmed meson systems~\cite{Braaten:2003he}, 
three-pion systems~\cite{Hyodo:2013zxa}, and hyperon systems such as the $nn\Lambda$~\cite{Ando:2015fsa}. 

In this work we have shed new light on the meson-nucleus systems having a strange (charm) quark, from the 
viewpoint of low-energy universality. We focus on the $\bar{K}NN$ ($DNN$) system with $J=0$, $I=3/2$, 
and $I_{3}=-3/2$, or more specifically, the $K^{-}nn$ ($D^{0}nn$) system. In these systems, all the two-body 
interactions occur in the $I=1$ combinations. In comparison to other quantum numbers, the quantum numbers 
associated with the $K^{-}nn$ ($D^{0}nn$) channel are ideal for examining the Efimov effect due to the following 
reasons. First, the absence of the Coulomb interaction guarantees that the low-energy behavior of this system 
is only governed by the two-body scattering lengths of the strong interaction. Second, the absence of any nearby 
coupled channels is also suitable for our purpose, otherwise the existence of such coupled channels is known 
to generally reduce the effective attraction in the three-body system~\cite{Braaten:2003he,Hyodo:2013zxa} 
diminishing the likelihood of formation of Efimov bound states.

While the scattering length of the two-neutron system is much larger than the typical length scale of the strong 
interaction, we should note that the magnitude of the physical $K^{-}n$ ($D^{0}n$) scattering length is not large enough for 
the meson-neutron two-body system to become resonant. However, our analysis reveals that the meson-neutron scattering 
length is expected to reach the unitary (resonant) limit, through an extrapolation of the quark mass from the strange sector
to the charm, where the couplings to sub-threshold decay channels are eliminated. Consequently with such a {\it zero coupling 
limit} (ZCL), the otherwise {\it complex} meson-neutron scattering lengths turn out to be {\it real valued}. This kind of
idealization obviously leads to a considerable simplification of the analysis of the three-body {\it Faddeev-like} integral 
equations compared to a more involved ``realistic'' analysis with complicated coupled-channel dynamics involving also 
hadrons other than the $K^-$ ($D^0$) meson and the neutron. Such a sophisticated approach is currently beyond the scope of 
our {\it naive} low-energy effective field theoretical (EFT) framework used in this work. Our primary objective in this 
paper is to present a simple leading order EFT analysis in an idealized limit to illuminate certain aspects of remnant two- 
and three-body universal physics that may be associated with bound state ({\it dimer} and {\it trimer}) formation in physical 
systems such as $K^{-}n$ ($D^{0}n$) and $K^{-}nn$ ($D^{0}nn$). In principle, the universal features of unphysical quark mass 
system can be studied in lattice QCD simulations, as previously demonstrated in the context of few-nucleon 
systems~\cite{Barnea:2013uqa}.

This paper is organized as follows. In Sec.~\ref{sec:twobody}, we study the two-body interaction of a flavored 
meson ($K^{-}$ or $D^{0}$) with a neutron in the $I=1$ state. We introduce a coupled-channel scattering model 
to perform an extrapolation from the strangeness sector to that of the charm. We evaluate the two-body scattering 
length in this extrapolation, together with the ZCL where the couplings to sub-threshold decay channels are artificially 
switched off. In Sec.~\ref{sec:threebody}, we formulate a {\it pionless} cluster effective field theory which 
describes the low-energy properties and dynamics of a three-particle cluster state consisting of two neutrons and 
a flavored meson ($K^-$ or $D^0$). In particular, we consider a possible scenario where the $s$-wave meson-neutron 
scattering length is infinitely large, while the $s$-wave neutron-neutron scattering length is fixed at 
its physical value. We perform an asymptotic analysis of the system of three-body integral equations to examine the 
renormalization group (RG) {\it limit cycle} behavior and other associated features related to the Efimov effect. 
Our numerical results for the full non-asymptotic analysis of the integral equations with and without including a 
three-body interaction are then presented. The final section is devoted to a summary of this work. Furthermore, 
based on our numerical results, a plausibility argument is presented at a qualitative level on the feasibility of 
$K^-nn$ or $D^0nn$ bound trimer formation. A discussion on certain numerical methodologies adopted in this work has 
been relegated to the appendices. 

\section{Two-body meson-neutron interaction}\label{sec:twobody}

\subsection{$K^{-}n$ system, $D^{0}n$ system, and the unitary limit}

We first summarize the known properties of the meson-neutron interactions. In the $\bar{K}N$ system, 
several experimental data constrain the meson-baryon scattering amplitude~\cite{Hyodo:2011ur}. 
Among others, the recent measurement of the kaonic hydrogen by the SIDDHARTA 
collaboration~\cite{Bazzi:2011zj,Bazzi:2012eq} gives a strong constraint on the low-energy $\bar{K}N$ 
interaction, because the result is directly related to the $K^{-}p$ scattering length through the 
improved Deser-type formula~\cite{Meissner:2004jr}. The analysis of the meson-baryon scattering with a 
complete {\it next-to-leading} order chiral SU(3) dynamics including the SIDDHARTA 
constraint~\cite{Ikeda:2011pi,Ikeda:2012au} determines the scattering lengths of the $K^{-}n$ system 
as\footnote{In this paper, the scattering length is defined as $a_{0}=-f(E=0)$ with the scattering amplitude 
$f$. The sign convention of the scattering length is opposite to that used in Refs.~\cite{Ikeda:2011pi,Ikeda:2012au}.}
\begin{align}
    a_{0,K^{-}n}
    &= 
    -0.57^{+0.21}_{-0.04}-i0.72_{-0.26}^{+0.41}\text{ fm} .
    \label{eq:Kmnslength}
\end{align}
The negative real part indicates that the $K^{-}n$ interaction is attractive, but not strong enough to 
support a quasi-bound state. The imaginary part of the scattering length indicates possible decay into 
sub-threshold $\pi\Sigma$ and $\pi\Lambda$ channels.

The $D^{0}n$ system is a counterpart of the $K^{-}n$ system in the charm sector. In contrast to the 
strangeness sector, there is no experimental data for the $D^{0}n$ scattering process. Theoretically, the $D^{0}n$ 
interaction has been studied by generalizing the established models in the 
strangeness sector~\cite{Hofmann:2005sw,Mizutani:2006vq,GarciaRecio:2008dp,Haidenbauer:2010ch,Yamaguchi:2011xb} 
(see Ref.~\cite{Hosaka:2016ypm} for a recent review). Here we take an alternative strategy of using the 
experimental information of the $\Sigma_{c}(2800)$ resonance. The mass and width of the neutral 
$\Sigma_{c}^{0}(2800)$ state are given by the Particle Data Group (PDG)~\cite{Olive:2016xmw} as
\begin{align}
    M_{\Sigma_{c}^{0}(2800)}
    &= 
    2806^{+5}_{-7}\text{ MeV} ,\quad
    \Gamma_{\Sigma_{c}^{0}(2800)}
    = 
    72^{+22}_{-15}\text{ MeV} ,
    \label{eq:Sigmac2800}
\end{align}
which lies very close to the $D^{0}n$ threshold at $\sim 2803.8$ MeV. Although the spin and parity of 
$\Sigma_{c}(2800)$ are not yet determined experimentally, by assuming $J^{P}=1/2^{-}$, we can determine 
the strength of the $s$-wave interaction in the $D^{0}n$ system. If the resonance pole of $\Sigma_{c}(2800)$ 
is found to lie below the $D^{0}n$ scattering threshold, then the $D^{0}n$ interaction is attractive enough 
to generate a quasi-bound state below the threshold. As will be demonstrated below, such a $D^{0}n$ quasi-bound 
picture of the $\Sigma_{c}(2800)$ resonance naturally arises in the SU(4) contact interaction model.

From these observations, we can draw the following conclusion regarding the meson-neutron interaction. On the 
one hand, the $K^{-}n$ system has a weakly attractive scattering length, as indicated by recent analysis in the 
strangeness sector. On the other hand, the $D^{0}n$ system can support a quasi-bound state below the threshold, 
which is ostensibly identified with the observed $\Sigma_{c}(2800)$ state. If we perform an extrapolation of the 
$K^{-}n$ interaction to the $D^{0}n$ interaction by changing the mass of the $s$-quark to that of the $c$-quark, 
we can expect the existence of an unphysical region of the quark mass between $m_{s}$ and $m_{c}$ where a very 
shallow quasi-bound state is formed when the magnitude of the scattering length becomes infinitely large. 
This represents the universal region around the {\it unitary limit} of the meson-neutron interaction. The 
situation is analogous to the two-nucleon~\cite{Braaten:2003eu}, the two-pion~\cite{Hyodo:2013zxa}, and the 
$\Lambda\Lambda$~\cite{Yamaguchi:2016kxa} systems, where the unitary limit of the two-hadron scattering is 
achieved by the quark mass extrapolation into the unphysical region. In much the same way we plan in this work
to explore the universal physics in proximity to meson-neutron unitarity using unphysical quark masses between 
the strangeness and charm limits. 

\subsection{Models of $\bar{K}N$ and $DN$ amplitudes}

To demonstrate that the unitary limit is indeed realized through the extrapolation, we employ a contact interaction 
model with flavor symmetry. In the strangeness sector, the {\it Weinberg-Tomozawa model} in the chiral SU(3) dynamics 
successfully describes the $\bar{K}N$ scattering~\cite{Kaiser:1995eg,Oset:1998it,Oller:2000fj,Hyodo:2011ur}. In the 
charm sector, the SU(4) generalization of this approach is developed in Ref.~\cite{Mizutani:2006vq} in which the 
$\Lambda_{c}(2595)$ resonance in the $DN$ scattering is dynamically generated in the $I=0$ sector, analogous to 
$\Lambda(1405)$ in the strangeness sector. Evidently, there is one model in each sector which successfully describes the known 
experimental data. Therefore, in this work we describe both $\bar{K}N$ and $DN$ systems within a unified framework of 
a {\it dynamical coupled-channel model}. This enables us to perform the flavor extrapolation from strangeness to charm.

The coupled-channel scattering amplitude $T_{ij}(W)$ with the total energy $W$ is given by the scattering 
equation
\begin{align}
    T(W)=[V^{-1}(W)-G(W)]^{-1} ,
\end{align}
with the interaction kernel $V(W)$:
\begin{align}
    V_{ij}(W) 
    =
    -\frac{C_{ij}}{f_{i}f_{j}}(2W-M_{i}-M_{j})
    \sqrt{\frac{M_{i}+E_{i}}{2M_{i}}}
    \sqrt{\frac{M_{j}+E_{j}}{2M_{j}}} ,
    \label{eq:contactterm}
\end{align}
where $C_{ij}$ is the coupling strength matrix specified below [see Eq.~\ref{eq:Cmatrix}], $M_{i}$ ($E_{i}$) is the 
mass (energy) of the baryon in channel $i$, and $f_{i}$ is the channel-dependent meson decay constant. The 
loop function $G_{i}(W)$ has a logarithmic ultraviolet divergence which we tame employing dimensional regularization. 
After removal of the pole term through regularization, the finite constant parts proportional to $\ln(4\pi)$, the Euler 
constant $\gamma_E$, etc., appearing in the loop function are then replaced by the {\it subtraction constant} $a_{i}(\mu_{\rm reg})$ 
at the regularization scale $\mu_{\rm reg}$. Such a term acts similar to a counter-term in a renormalizable 
theory (see, e.g., Refs.~\cite{Oller:2000fj,Hyodo:2008xr}). 
Thus, we have
\begin{align}
    G_{i}(W;a_{i})
    &=
    \frac{2M_{i}}{16\pi^{2}}
    \Bigl\{
    a_{i}(\mu_{\rm reg})+\ln\frac{m_{i}M_{i}}{\mu^{2}_{\rm reg}}
    +\frac{M_{i}^{2}-m_{i}^{2}}{2W^{2}}\ln\frac{M_{i}^{2}}{m_{i}^{2}}
    \nonumber \\
    &\quad +\frac{\bar{q}_{i}}{W}
    [\ln(W^{2}-(M_{i}^{2}-m_{i}^{2})+2W\bar{q}_{i}) \nonumber \\
    &\quad +\ln(W^{2}+(M_{i}^{2}-m_{i}^{2})+2W\bar{q}_{i}) \nonumber \\
    &\quad -\ln(-W^{2}+(M_{i}^{2}-m_{i}^{2})+2W\bar{q}_{i}) \nonumber \\
    &\quad -\ln(-W^{2}-(M_{i}^{2}-m_{i}^{2})+2W\bar{q}_{i})]
    \Bigr\} ,
    \label{eq:loop}
\end{align}
where $m_{i}$ denotes the mass of the meson in the channel $i$, and the quantity 
\begin{equation}
\bar{q}_{i}=\sqrt{[W^{2}-(M_{i}-m_{i})^{2}][W^{2}-(M_{i}+m_{i})^{2}]}/(2W) 
\end{equation}
is the analytically continued three-momentum in the center-of-mass frame. In this model, the interaction strengths 
$C_{ij}$ are basically determined by the flavor symmetry. The free parameters in the model are the subtraction constants 
$a_{i}$. These real valued subtraction constants play the role of ultraviolet cutoffs, and effectively renormalize the 
dynamics of the channels which are not explicitly included in the model space. In this convention, the (diagonal) scattering 
length $a_{0,i}$ in channel $i$ is given by
\begin{align}
    a_{0,i}
    &=
    \frac{M_{i}}{4\pi (M_{i}+m_{i})}T_{ii}(M_{i}+m_{i})\,.
\end{align}

Next we consider the model space of the scattering. Because we are interested in the energy region 
near the $\bar{K}N/DN$ threshold, the channels with higher energies than the $\bar{K}N/DN$ thresholds 
(such as $\eta\Sigma/\eta\Sigma_{c}$ and $K\Xi/K\Xi_{c}$) are not very relevant to our analysis which we 
exclude.\footnote{
By fitting the subtraction constants to experimental data, the contribution from the higher energy channels 
can be effectively renormalized. In fact, it is phenomenologically shown in Ref.~\cite{Hyodo:2007jq} 
that the effect of the higher energy channels is not very strong around $\bar{K}N$ threshold energy region.} 
On the other hand, we include all relevant channels lower than the $\bar{K}N$ and $DN$ thresholds, respectively,
namely,
\begin{align}
    K^{-}n,\quad \pi^{-}\Lambda,\quad \pi^{0}\Sigma^{-},\quad \pi^{-}\Sigma^{0}
\label{eq:strangemodelspace}
\end{align}
in the strangeness sector, and the corresponding open channels in the charm sector:
\begin{align}
    D^{0}n,\quad \pi^{-}\Lambda_{c}^{+},
\quad \pi^{0}\Sigma_{c}^{0},\quad \pi^{-}\Sigma_{c}^{+}\,.
\end{align}
Here the coupling matrix for these channels takes the form
\begin{align}
    C
    =
    \begin{pmatrix}
    1 & -\sqrt{\frac{3}{2}}\kappa 
      & -\sqrt{\frac{1}{2}}\kappa & \sqrt{\frac{1}{2}}\kappa \\
      & 0 & 0 & 0 \\
      &  & 0 & -2 \\
      &  &   & 0 \\
    \end{pmatrix} .
\label{eq:Cmatrix}
\end{align}
A comment on the suppression factor $\kappa$ is in order. As discussed in Ref.~\cite{Mizutani:2006vq}, the microscopic 
origin of the contact interaction, Eq.~\eqref{eq:contactterm}, can be understood in terms of vector meson exchange picture. 
In this case, the interaction strength is proportional to $1/m_{V}^{2}$ with $m_{V}$ being the mass of the exchanged vector 
meson. While SU(4) symmetry exactly requires $\kappa=1$, for the heavy flavor exchange processes, $C_{1i}$ ($i=2,3,4$) should 
be significantly suppressed. Thus, as in Ref.~\cite{Mizutani:2006vq}, the suppression factor in the charm sector, $\kappa<1$, 
is introduced to account for the SU(4) breaking effects of the underlying mechanism. The masses of hadrons relevant for 
these channels are taken from the central values provided by the PDG~\cite{Olive:2016xmw}, and for for 
convenience summarized in Table~\ref{tbl:hadronmass}. We use the physical meson decay constants, $f_{\pi}=92.4$ MeV and 
$f_{K}=109.0$ MeV, and the decay constant of the $D$ meson is also chosen to be $f_{D}=92.4$ MeV, following 
Ref.~\cite{Mizutani:2006vq}. The regularization scale is set at $\mu_{\rm reg}=1$ GeV, consistent with the typical scale 
of the vector mesons not explicitly incorporated in this framework.\footnote{
It is important to note that the regularization scale can be chosen arbitrarily to suite the purpose of a given theory or 
model as long as the physical consequences are independent of this scale. This is, however, not to be confused with the hard 
or the {\it breakdown} scale of the same theory which in our context may be chosen as the pion mass, i.e., 
${}^{\pi\!\!\!/}\Lambda\sim m_\pi$, in commensurate with the {\it pionless} EFT analysis pursued in the three-body sector.} 

\begin{table*}[tbp]
\caption{Masses of hadrons.}
\begin{ruledtabular}
\begin{tabular}{llllllllllll}
Hadron & $K^{-}$ & $\pi^{-}$ & $\pi^{0}$ & $n$ & $\Lambda$ & $\Sigma^{-}$ & $\Sigma^{0}$ & $D^{0}$ & $\Lambda_{c}^+$ & $\Sigma_{c}^{0}$ & $\Sigma_{c}^{+}$ \\
\hline
Mass [MeV] & $493.677$ & $139.57018$ & $134.9766$ & $939.565379$ & $1115.683$ & $1197.449$ & $1192.642$ & $1864.3$ & $2286.46$ & $2453.75$ & $2452.9$ \\
\end{tabular}
\end{ruledtabular}
\label{tbl:hadronmass}
\end{table*}%

The remaining parameters are the subtraction constant $a_{i}$ in each channel and the suppression factor 
$\kappa$ in the coupling strength matrix, appearing in Eq.~\eqref{eq:Cmatrix}. To determine these parameters 
in the strangeness sector ($a_{i}^{s}$ and $\kappa^{s}$), it is essential to take into account the SIDDHARTA data 
of the kaonic hydrogen~\cite{Bazzi:2011zj,Bazzi:2012eq}. The data allows a direct extraction of the $K^{-}p$ 
scattering length as related through the improved Deser-type formula~\cite{Meissner:2004jr} and provides a strong 
constraint on the low-energy $\bar{K}N$ interaction~\cite{Ikeda:2011pi,Ikeda:2012au}. In Ref.~\cite{Ikeda:2012au}, 
a simplified model (termed as the ``ETW'' model) is constructed with the Weinberg-Tomozawa interaction ($\kappa_{s}=1$) 
in the model space of Eq.~\eqref{eq:strangemodelspace}, but quite reasonably reproducing the full next-to-leading 
order amplitude constrained by the SIDDHARTA data. The explicit values of the subtraction constants $a^{s}_{i}$ 
in the ETW model are summarized in Table~\ref{tbl:subtraction}. The scattering length of the $K^{-}n$ system is 
obtained as
\begin{align}
    a_{0,K^{-}n}
    &= 
    -0.135-i0.410\text{ fm} .
\label{eq:a_Kn_fullmodel}
\end{align}
Although the value deviates from the full result in Eq.~\eqref{eq:Kmnslength} due to the simplified assumptions of 
this scattering model, both the order of magnitude and the qualitative features of the result (e.g., weak attraction) 
remain unchanged. We think this is sufficient for the purpose of our present analysis.\footnote{
We have checked that the central value given in Eq.~\eqref{eq:Kmnslength} can be reproduced by adjusting 
$a_{1}^{s}=-0.649$ and $a_{2}^{s}=-1.899$. The qualitative conclusion in Sec. II, i.e., the emergence of the unitary 
limit between strangeness and charm sectors in the ZCL, remains unchanged with this parameter set. In this paper, we 
use the original subtraction constants of the ETW model~\cite{Ikeda:2012au}, respecting the consistency with 
the $K^{-}p$ sector.}

\begin{table}[tbp]
\caption{Subtraction constants $a_{i}^{s,c}(\mu_{\rm reg})$ at the regularization scale $\mu_{\rm reg}=1$ GeV.}
\begin{ruledtabular}
\begin{tabular}{lcccc}
channel $i$ & 1 & 2 & 3 & 4 \\
\hline
$a^{s}_{i}$ (strangeness) & $-1.283$ & $0.238$ & $-0.714$ & $-0.714$ \\
$a^{c}_{i}$ (charm) & $-1.663$ & $-2.060$ & $-2.060$ & $-2.060$ \\
\end{tabular}
\end{ruledtabular}
\label{tbl:subtraction}
\end{table}%

In the charm sector, the subtraction constants $a^{c}_{i}$ were determined in Ref.~\cite{Mizutani:2006vq} by  
the argument of ``natural size''~\cite{Oller:2000fj,Hyodo:2008xr}. However, a slight modification of the constant 
associated with the $DN$ channel was necessary to reproduce the observed $\Lambda^{+}_{c}(2595)$ resonance in the 
$I=0$ channel amplitude. Likewise, the same strategy is pursued in the present work for the $I=1$ two-body 
subsystem; the subtraction constants in channels $i=2 - 4$ (i.e., $\pi^{-}\Lambda_{c}^+,\,\pi^{0}\Sigma_{c}^{0}$ 
and $\pi^{-}\Sigma_{c}^{+}$) are chosen to be $-2.060$~\cite{Bayar:2012dd}, consistent with the ``natural size'' 
$a_{\rm nat}\sim -2$~\cite{Oller:2000fj,Hyodo:2008xr},\footnote{
There are two arguments determining the natural size of the subtraction constants. In Ref.~\cite{Oller:2000fj}, 
through a matching of the loop function evaluated using dimensional and cut-off regularization methods, the natural 
size of these constants were estimated. In Ref.~\cite{Hyodo:2008xr}, the determination of the natural values of these 
constants was pursued utilizing the renormalization condition, $G_{i}(W=M_{i};a_{\rm nat})=0$, along with the constraints from 
the low-energy behavior of the amplitudes and the physical requirements on the loop functions. In both cases, the 
natural value is estimated as $a_{\rm nat}\sim -2$ at the regularization scale, $\mu_{\rm reg}=1$ GeV.} 
while the subtraction constant in channel 1 (i.e., for $D^{0}n$) is slightly adjusted to reproduce the $\Sigma_{c}(2800)$ 
state near the $D^{0}n$ threshold. Such fine-tuning of the value of the subtraction constant reflects further SU(4) breaking 
effects in our model. We also tune the value of $\kappa^{c}$ to control the width of the resonance. By choosing the subtraction 
constants $a^{c}_{i}$ shown in Table~\ref{tbl:subtraction} with $\kappa^{c}=0.453$, we dynamically generate a resonance pole 
corresponding to the $\Sigma_{c}(2800)$ state at
\begin{align}
    M
    &= 
    2800\text{ MeV} ,\quad
    \Gamma
    = 
    72\text{ MeV} .
\end{align}
In this model, the $D^{0}n$ scattering length is found to be
\begin{align}
    a_{0,D^{0}n}
    &= 
    0.764-i0.615\text{ fm} .
\label{eq:a_D0n_fullmodel}
\end{align}
The positive real part is in accordance with the existence of the quasi-bound state $\Sigma_{c}(2800)$ 
below the threshold. The origin of the $I=1$ quasi-bound state can be understood in the following argument. 
Because of the SU(4) symmetry, the interaction of the $D^{0}n$ channel has the same sign as that of 
the $K^{-}n$ channel, while the strength at the threshold is enhanced by the ratio $m_{D}/m_{K}$. Thus, we 
can expect a stronger attractive interaction in the $D^{0}n$ channel. Moreover, the heavier reduced mass 
of the $DN$ system leads to suppression of the kinetic energy which is implicitly reflected by the larger 
value of the two-body scattering length. This is a crucial expedient in the manifestation of universality 
in the three-body $D^0nn$ system as demonstrated in the next section. 

\subsection{Flavor extrapolation and zero coupling limit}\label{subsec:extrapolation}

Next we introduce a parameter $0\leq x\leq 1$ which controls the extrapolation from strangeness 
to charm. Assuming a linear relationship among the following model parameters, we vary each as a 
function of $x$;
\begin{align}
    m_{i}(x)&=m_{i}^{s}(1-x)+m_{i}^{c} x , \\
    M_{i}(x)&=M_{i}^{s}(1-x)+M_{i}^{c} x , \\
    a_{i}(x)&=a_{i}^{s}(1-x)+a_{i}^{c} x , \\
    f_{i}(x)&= f_{i}^{s}(1-x) + f_{i}^{c}x ,\\
    \kappa(x)&= \kappa^{s}(1-x) + \kappa^{c}x ,
\end{align}
for the respective channels, $i=1,\dots 4$. Here, $x=0$ ($x=1$) corresponds to the 
physical point of the model in the strangeness (charm) sector, while all other intermediate 
values of $x$ represent the model in the unphysical domain. Thus, for instance, in the $i=1$ 
channel, by varying $x$ from 0 to 1, we can perform a linear extrapolation from the $K^{-}n$ 
to the $D^{0}n$ scattering sectors.

\begin{figure}[tbp]
    \centering
    \includegraphics[width=\columnwidth,bb=0 0 362 218]{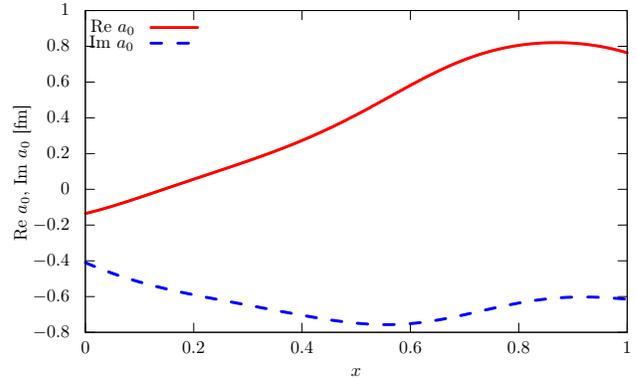}
\caption{\label{fig:slengthcomp}
Complex meson-neutron scattering length in the coupled-channel contact interaction model as 
a function of the extrapolation parameter $x$. Solid (dashed) line represents the real 
(imaginary) part.}
\end{figure}

\begin{figure}[tbp]
    \centering
    \includegraphics[width=\columnwidth,bb=0 0 362 218]{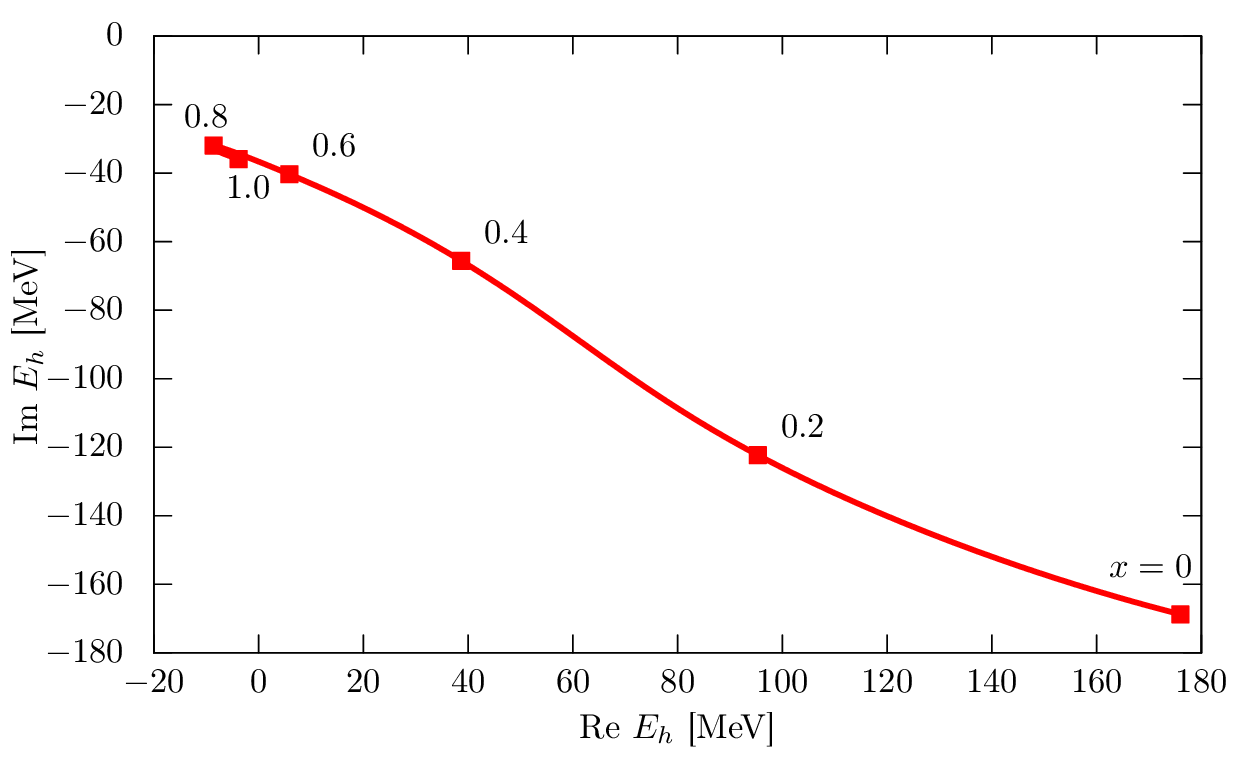}
\caption{\label{fig:trajectory_y1}
Trajectory of the pole of the scattering amplitude in the coupled-channel contact interaction 
model. The energy is measured with respect to the threshold of channel 1, i.e., 
$E_{h}=W-M_{1}(x)-m_{1}(x)$.}
\end{figure}

Now we study the behavior of the complex scattering length in the coupled-channel contact interaction 
model. The real and imaginary parts of the meson-neutron scattering length as functions of $x$ are shown 
in Fig.~\ref{fig:slengthcomp}. The scattering length varies continuously from $a_{0,K^{-}n}$ to $a_{0,D^{0}n}$ 
along with $x$. It is not immediately clear from this figure how to decipher the remnant of the universal 
features of meson-neutron unitarity in any straightforward manner due to effects of the sub-threshold decay 
channels. We note that in the present case, the $S$-matrix pole corresponding to the quasi-bound state of 
the $D^{0}n$ system moves to the higher positive energy region as we decrease $x$ from 1, eventually going 
far above the $K^{-}n$ threshold at $x=0$ without yielding a quasi-bound $K^{-}n$ state. The pole trajectory 
of the scattering amplitude is shown in Fig.~\ref{fig:trajectory_y1}. Note that the pole is on the Riemann 
sheet with physical momentum with respect to channel 1 and unphysical momenta in regard to the 
others.\footnote{
In an $n$-channel problem, the scattering amplitude is defined on a $2^{n}$-sheeted Riemann surface in the 
complex energy plane with momentum $p_{i}$ and $-p_{i}$ in channel $i$ corresponding to the same energy. The 
Riemann sheet is specified by choosing either the physical momentum ($\im p_{i}>0$, first sheet) or the 
unphysical momentum ($\im p_{i}<0$, unphysical sheet) for each channel $i$. The most adjacent Riemann sheet 
to the real scattering axis is obtained by choosing physical momenta in the open channels and unphysical 
momenta in the closed channels.} 
Thus, the corresponding energy pole, $E_h=W-M_{1}(x)-m_{1}(x)$ (measured with respect to the threshold of the 
channel 1), directly influences the physical spectrum when $\re E_{h}<0$ (as in the charm sector, $x=1$), being 
on the most adjacent Riemann sheet to the physical real axis. On the other hand, its effect becomes less 
significant when $\re E_{h}>0$ (as in the strangeness sector $x=0$), with the proximity of the pole position 
far away from the physical axis.

\begin{figure}[tbp]
    \centering
    \includegraphics[width=\columnwidth,bb=0 0 362 218]{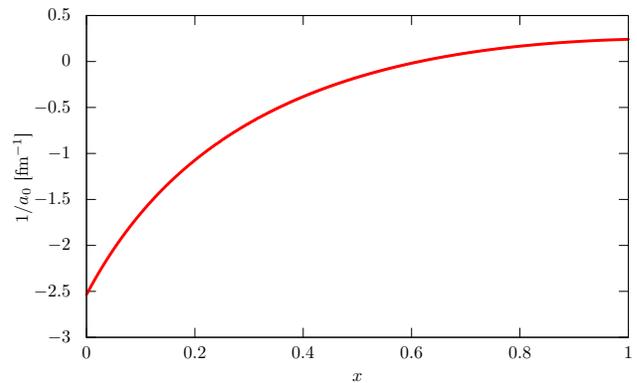}
\caption{\label{fig:slengthreal}
Meson-neutron inverse scattering length in the absence of decay channels in the zero coupling limit 
as a function of the flavor extrapolation parameter $x$.}
\end{figure}

\begin{figure}[tbp]
    \centering
    \includegraphics[width=\columnwidth,bb=0 0 362 218]{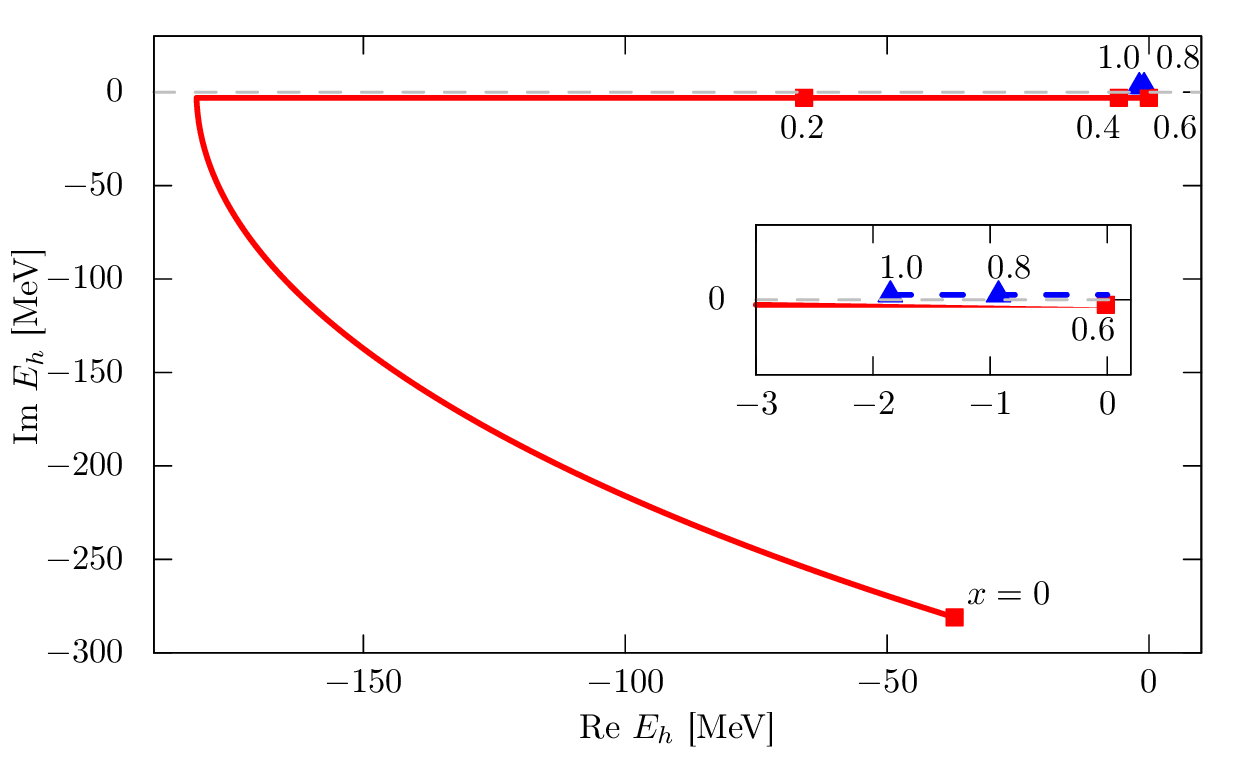}
\caption{\label{fig:trajectory_y0}
Trajectory of the pole of the scattering amplitude in the zero coupling limit. The energy is measured 
with respect to the threshold of channel 1 [$E_{h}=W-M_{1}(x)-m_{1}(x)$]. Dashed (solid) line represents 
the pole on the physical (unphysical) Riemann sheet.}
\end{figure}

To elucidate a possible scenario to access the unitary limit of the meson-neutron interaction, 
we consider a zero coupling limit (ZCL) in which the channel couplings are artificially switched 
off~\cite{Eden:1964zz,Pearce:1988rk,Cieply:2016jby}, i.e.,
\begin{align}
    C_{1i}&=C_{i1}=0 \quad \text{for } i=2,3,4 .
\end{align}
Under this assumption, the coupled-channel problem reduces to a single-channel scattering 
of the $K^{-}n/D^{0}n$ system. The $K^{-}n$ system is again found to have no bound state, 
while the previous quasi-bound state in the $D^{0}n$ channel now becomes a real bound state 
with invariant mass of $W\sim 2802$ MeV. Correspondingly, the scattering length is real 
and negative (positive) in the $K^{-}n$ ($D^{0}n$) channel:
\begin{align}
    a_{0,K^{-}n}^{\rm ZCL}
    &= 
    -0.394\text{ fm} ,\quad
    a_{0,D^{0}n}^{\rm ZCL}
    = 
    4.141\text{ fm} .
\label{eq:ZCL_ad}
\end{align}
The above numerical values of the scattering lengths in the ZCL strongly suggest that 
the meson-neutron interactions become resonant at an intermediate point in the unphysical 
domain ($0<x<1$). In Fig.~\ref{fig:slengthreal}, we display the result for the inverse 
scattering length $1/a_{0}(x)$, expressed as a function of the extrapolation parameter $x$. We 
indeed find that the unitary limit ($1/a_{0}= 0$) is achieved at 
$x= 0.615$. 
At this point, 
the extrapolated mass of the flavored meson in the channel 1 is
\begin{align}
    m_{1}(0.615)&=1336.61 \text{ MeV} .
\end{align}
The corresponding behavior of the pole trajectory in the zero coupling limit is shown in 
Fig.~\ref{fig:trajectory_y0}. The {\it real} bound state pole at $x=1$ that lies on the 
{\it physical (first) Riemann sheet} turns into a {\it virtual} state pole at the unitary 
limit moving onto the {\it unphysical Riemann sheet}. As we further decrease $x$, 
the virtual state pole eventually meets with a second virtual state pole on the unphysical 
sheet and then acquires a finite width~\cite{Hyodo:2014bda}. In particular, at $x=0$, the 
virtual state pole remains still on the unphysical sheet with a finite imaginary part
below the threshold.

As seen earlier the neglect of the decay channels in the ZCL yields real scattering lengths
which may appear somewhat unrealistic in contrast to the full coupled-channel model that yields 
real and complex parts of the scattering lengths of roughly the same size. A natural question 
therefore arises as to the utility of the ZCL approach. To this end, let us make a comment to elucidate
the significance of the ZCL analysis. One may think that the behavior of the scattering length 
(Fig.~\ref{fig:slengthreal}) and the pole trajectory (Fig.~\ref{fig:trajectory_y0}) in the ZCL are 
very much different from those in the full model (Figs.~\ref{fig:slengthcomp} and \ref{fig:trajectory_y1}). 
However, the qualitative feature, namely, the existence of the pole remains unchanged by taking the ZCL. 
In fact, it is guaranteed by a topological invariant of the scattering amplitude that the existence of a 
pole is stable against the continuous change (deformation) of the model parameters, except for the special 
case where a zero of the amplitude exists near the pole (vide Ref.~\cite{Kamiya:2017pcq} for details). 
Because the ZCL can be achieved by gradually decreasing the off-diagonal couplings $C_{1i}=C_{i1}$, it is 
{\it continuously connected} with the full model. Hence, it is worth studying the model in the ZCL to check 
whether the system develops an eigenstate pole, although the exact position of the pole may show a large 
deviation. Furthermore, as we will show in the next subsection, the relation between the scattering lengths 
and the pole positions exhibits two-body universality, both in the full model and the model in the ZCL. Thus, 
the ZCL model and the full model share common features of universal physics in the two-body sector. 

\subsection{Universality in the two-body sub-system}

It is instructive to take another look at the above results from the viewpoint of two-body 
universality. When the magnitude of the scattering length $|a_{0}|$ becomes large and approaches 
the unitary limit, universality suggests that there exists a two-body meson-neutron eigenstate 
with the eigenmomentum
\begin{align}
    k_{a}= i/a_{0} ,
\end{align}
up to corrections suppressed by effective range terms. This relation holds even for a complex value 
of $a_{0}$. Thus, if we calculate the eigenmomentum $k_{h}$, that corresponds to the pole position of 
the scattering amplitude in the system with a large scattering length, the true eigenvalue $k_{h}$ 
can be well approximated by $k_{a}$ in the two-body universal region. The deviation of $k_{h}$ from 
$k_{a}$ thus serves as a measure of the violation of universality. 

\begin{figure}[tbp]
    \centering
    \includegraphics[width=\columnwidth,bb=0 0 362 218]{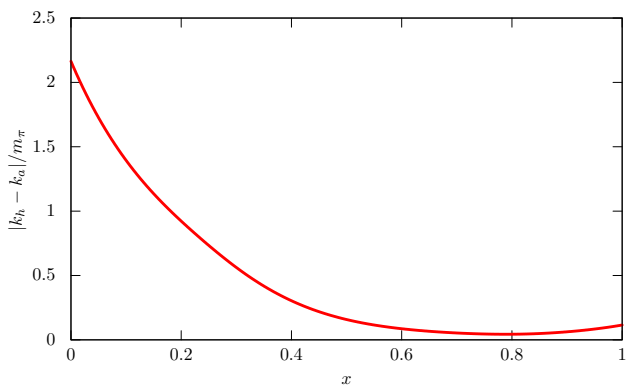}
    \caption{\label{fig:eigenmomentumcomp}
Deviation of the eigenmomentum from the universal prediction $|k_{h}-k_{a}|/m_{\pi}$ as a function of 
the flavor extrapolation parameter $x$ in the coupled-channel contact interaction model.}
\end{figure}

We show in Fig.~\ref{fig:eigenmomentumcomp} the deviation of the eigenmomentum from the universal 
prediction $|k_{h}-k_{a}|$ normalized by the typical momentum scale of the strong interaction, i.e., 
the pion mass $m_{\pi}$, as a function of $x$. Although there is a large deviation in the strangeness 
sector ($x=0$), the universal value $k_{a}$ around $x\sim 0.8$, including the $D^{0}n$ system at $x=1$, 
gives a reasonable prediction of the true eigenvalue $k_{h}$. Since the deviation is much smaller than 
the typical scale $m_\pi$, the result suggests that the full coupled-channel model, despite the influence 
of the decay channels, should also reflect the relevance of low-energy universality in the meson-neutron 
system close to the charm quark sector, $x\sim 1$.

\begin{figure}[tbp]
    \centering
    \includegraphics[width=\columnwidth,bb=0 0 362 218]{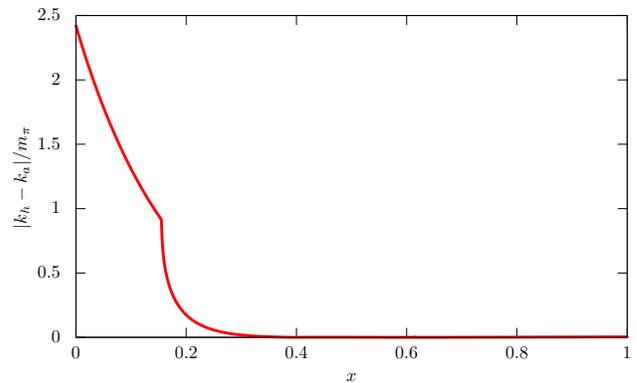}
\caption{\label{fig:eigenmomentumreal}
Deviation of the eigenmomentum from the universal prediction $|k_{h}-k_{a}|/m_{\pi}$ in the zero coupling limit as 
a function of the flavor extrapolation parameter $x$.}
\end{figure}

The situation becomes much more conspicuous in the ZCL. As shown in Fig.~\ref{fig:eigenmomentumreal}, 
we find quite a good agreement of $k_{h}$ with $k_{a}$, not only in the vicinity of the unitary limit 
($x\sim 0.6$) but also in the entire region $x\gtrsim 0.3$. We should note that the eigenstate 
in the unbound region ($x\lesssim 0.6$) corresponds to a virtual state ($a_0<0$), an unphysical 
bound state with the S-matrix pole lying on the unphysical  sheet. A characteristic cusp-like structure 
around $x\sim 0.15$ is caused when the virtual state acquires a finite width (also see Fig.~\ref{fig:trajectory_y0}). 

Thus, our simplistic model approach demonstrates that universality is a powerful tool to investigate the 
meson-neutron two-body system, both in the full model and the ZCL. 

\section{Three-body system of two neutrons and one meson}\label{sec:threebody}

\subsection{Effective field theory}

In this section, we consider the universal features in the three-body system consisting of two neutrons 
and one flavored meson with $J=0$, $I=3/2$, and $I_{3}=-3/2$, assuming that the two-body physics is {\it finely tuned} 
and characterized only by the two-body scattering lengths. A low-energy EFT can provide a simple but powerful 
systematic tool, for a quantitative analysis of such three-body systems in proximity to the scattering threshold. 
Recently, plethora of EFT based analyses have been used to predict the formation of bound states in a variety of 
three-body cluster states~\cite{Ando:2015fsa,Hammer:2001ng,AB10b,AYO13,AO14,Hammer:2017tjm} which are especially 
neutron-rich.       

Primarily for the sake of a qualitative exploratory study of three-body universality, we henceforth neglect 
the sub-threshold decay channels $i=2-4$, which means that our three-body analysis is consistent with the ZCL 
idealization with real valued meson-neutron scattering lengths [see Eq.~(\ref{eq:ZCL_ad})]. 
Our motivation of this part of the study is to determine under what circumstance do we expect to find 
the three-body system to become bound solely by virtue of the Efimov ``attraction''. 
There is currently no consensus, either experimentally or theoretically, as to whether the system in the 
physical limits, $x=0$ (i.e., $K^-nn$) and $x=1$ ($D^0nn$), respectively, are bound. Through universality based 
arguments we hope to gain general insights alternative to rigorous realistic calculations which is beyond the 
scope of the present work.      

Here we employ a leading order cluster effective field theory with a flavored meson $K^-$ ($D^0$) and a neutron 
$\psi_n$ as the elementary fields in the theory. We mainly focus on low-energy threshold states far below the pion 
mass, i.e., the theory is {\it pionless}, with explicit pion degrees of freedom and their interaction effectively 
integrated out. We note that the {\it breakdown scale} of the pionless EFT (i.e., ${}^{\pi\!\!\!/}\Lambda\sim m_\pi$) 
is different from that in Sec.~\ref{sec:twobody}. To establish the flavor extrapolation of the meson-neutron 
scattering lengths we have explicitly included pions in the SU(4) model with vector meson exchange. Once this is 
settled, we concentrate on the states near the threshold by utilizing the pionless theory.  

A power counting rule~\cite{Kaplan:1998tg,Kaplan:1998we,vanKolck:1998bw} for two-body contact interactions for 
such finely-tuned systems, incorporated with the {\it power divergence subtraction} scheme, is most suitable for 
renormalizing the strongly interacting two-body sector. In this section, we display the effective Lagrangian of 
the three-body system, where we denote the generic flavored meson field as $K$ for brevity, but what it basically 
stands for is the flavor extrapolated meson field $K(x)$, implicitly dependent on the parameter $0\leq x\leq 1$. 
Note here that the actual meson charge is irrelevant in our context without Coulomb interaction, so that the 
change in the charge from the antikaon $K^-$ (for $x=0$) to the $D^0$ meson (for $x=1$) should not be a matter 
of concern. 

The non-relativistic effective Lagrangian for the three-body system, consistent with the usual low-energy symmetries, 
like parity invariance, charge conjugation, time-reversal invariance, and small velocity Lorentz invariance, can 
be constructed as  
\begin{align}
    \mathcal{L}
    &= 
    \mathcal{L}_{K}
    +\mathcal{L}_{n}
    +\mathcal{L}_{\text{2-body}}
    +\mathcal{L}_{\text{3-body}} .
\end{align}
As regards the fundamental fields, the neutron is represented by a two-component (spin-doublet) spinor $\psi_{n}$, 
and the flavor extrapolated meson is represented by a spin-singlet field, $K(x)$. The standard forms of the single 
``heavy'' particle Lagrangians, $\mathcal{L}_{K}$ and $\mathcal{L}_{n}$, at the leading order are well 
known~\cite{Braaten:2004rn,BS01,AH04,Ando:2015fsa}, and their expressions will not be repeated here.
However, in the two-body sector, we explicitly spell out the relevant interaction parts of the leading order 
Lagrangian $\mathcal{L}_{\text{2-body}}$, projecting the dominant $s$-wave terms. When the $nn$ and flavor extrapolated 
$nK$ scattering lengths are much larger than the typical length scales of short-range interactions, the two-body 
interactions can be expressed by means of contact terms with couplings $g_{nK}$ and $g_{nn}$, parametrizing the 
short-distance UV physics:
\begin{align}
    \mathcal{L}^{int}_{\text{2-body}}
    &= -g_{nn}
    \left(\psi_{n}^{T}P_{(nn)}^{({}^{1}{\rm S}_{0})}\psi_{n}
    \right)^{\dag}
    \left(\psi_{n}^{T}P_{(nn)}^{({}^{1}{\rm S}_{0})}\psi_{n}
    \right) \nonumber \\
    &\quad -
    g_{nK}\phi_{K}^{\dag}\psi_{n}^{\dag}
    \psi_{n}\phi_{K}+\dotsb ,
    \label{eq:2body}
\end{align}
where the spin-singlet projection operator is defined as 
\begin{align}
    P_{(nn)}^{({}^{1}{\rm S}_{0})}
    &= 
    -\frac{i}{2}\sigma_{2} .
    \label{eq:nnprojection}
\end{align}
In this context, it is convenient to describe the three-body system in terms of auxiliary {\it diatom} or 
{\it dihadron}  fields~\cite{Braaten:2004rn,BS01,AH04}, namely, a spin-singlet $nn$-{\it dibaryon} field $s_{(nn)}$, 
and a spin-doublet $nK$-{\it dihadron} field $d_{(nK)}$. Through a Gaussian integration followed by a redefinition of 
the fields, the contact terms in the above two-body Lagrangian can be re-expressed as  
\begin{align}
    \mathcal{L}_{d(nK)}
    &= \frac{1}{g_{nK}}d^{\dag}_{(nK)}d_{(nK)}
    \nonumber \\
    &\quad 
    -\left[d_{(nK)}^{\dag}\psi_{n}\phi_{K}+\text{h.c.}\right]+\dotsb\,,\\
    \mathcal{L}_{s(nn)}
    &= \frac{1}{g_{nn}}s_{(nn)}^{\dag}s_{(nn)} 
    \nonumber \\
    &\quad 
    -\left[s_{(nn)}^{\dag}\left(\psi_{n}^{T}P_{(nn)}^{({}^{1}{\rm S}_{0})}\psi_{n}\right)+\text{h.c.}\right]+\dotsb \,,
\end{align}
where the ellipses in the above expressions denote the sub-leading interaction terms irrelevant 
in the context of our analysis.

Our three-body system exhibits the Efimov effect in the ZCL, as a straightforward consequence of the unitarity of the 
two-body amplitude, which is also demonstrated by the numerical analysis below. Following the basic EFT tenet, this 
necessitates the additional inclusion of a leading order non-derivatively coupled three-body contact term in the 
Lagrangian for renormalization (see, e.g., Refs.~\cite{Braaten:2004rn,BS01,AH04,BHV98} for details). 
In our case, such a three-body term is given by
\begin{align}
    \mathcal{L}_{\text{3-body}} &={\mathcal L}^{({}^{1}{\rm S}_{0})}_{nd(nK)}+\cdots\,,\nonumber \\
    {\mathcal L}^{({}^{1}{\rm S}_{0})}_{nd(nK)} &=
    -\frac{m_{K} g_{s}(\mu)}{\mu^{2}}
    \left(d^{T}_{(nK)}P_{(nd)}^{({}^{1}{\rm S}_{0})}\psi_{n}\right)^{\dag} \nonumber \\
    & \hspace{2cm} \!\! \times \left(d^{T}_{(nK)}P_{(nd)}^{({}^{1}{\rm S}_{0})}\psi_{n}\right)+\cdots ,
\label{eq:L3body}
\end{align}
with the spin-singlet projection operator
\begin{align}
    P_{(nd)}^{({}^{1}{\rm S}_{0})}
    &= 
    -\frac{i}{\sqrt{2}}\sigma_{2}\,,
\end{align}
and $g_{s}(\mu)$, an {\it a priori} unknown scale dependent three-body coupling. 
Here we especially choose to promote to the leading order, only the three-body interaction corresponding 
to the spin-singlet elastic channel, $nd_{(nK)}\to nd_{(nK)}$. The ellipses denote other allowed three-body 
interaction terms to be treated generally as sub-leading. It is important to note that precisely the presence of 
the asymptotic limit cycle in our case gives us the freedom to choose any one of the channels to promote the 
corresponding three-body term. 

\subsection{Three-body integral equation}

\begin{figure*}[tbp]
    \centering
    \includegraphics[width=17cm]{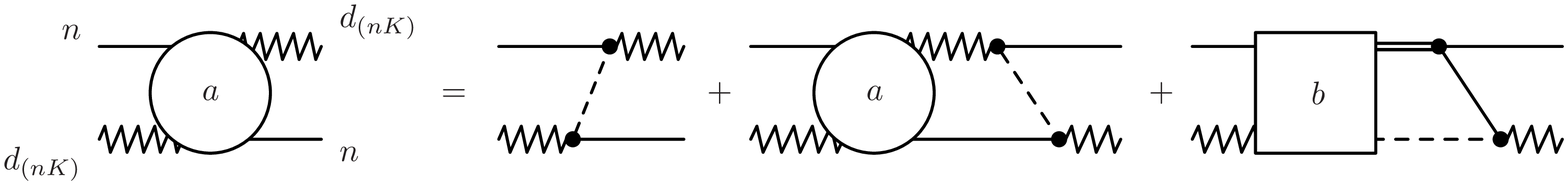}
    \includegraphics[width=12cm]{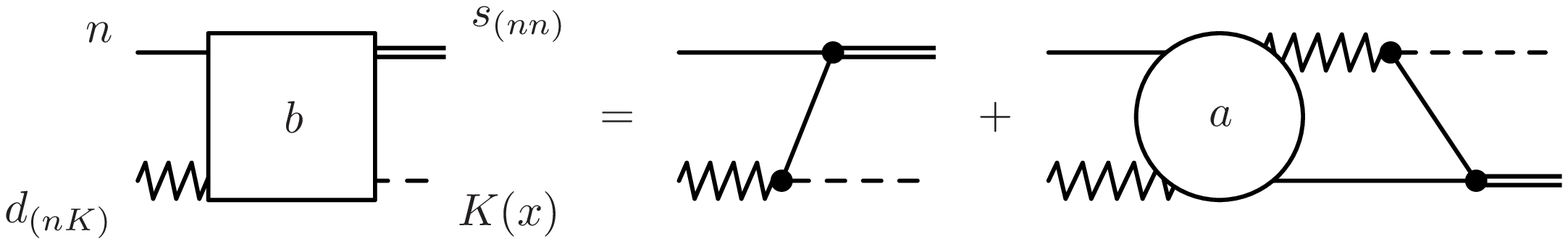}
\caption{\label{fig:diagram}
Feynman diagrams for the three-body coupled integral equations where the three-body contact terms are 
omitted for brevity. The solid and dashed lines represent the bare propagators of $n$ and $K(x)$, 
respectively, and the double and zigzag lines stand for the dressed dihadron propagators of $s_{(nn)}$ 
and $d_{(nK)}$, respectively.}
\end{figure*}

We now consider the integral equations for the meson-neutron-neutron three-body system in the total 
spin-singlet channel. Such a system of equations can be constructed either by combining the dibaryon field 
$s_{(nn)}$ and the flavored meson field $K$ or by combining the dihadron field $d_{(nK)}$ and the neutron 
field $n$. The system of equations consists of two kinds of Faddeev-like partitions: a direct (elastic) hadron 
exchange channel $nd_{(nK)}\to nd_{(nK)}$ (denoted as $t_{a}$), and a hadron rearrangement (inelastic) channel 
$nd_{(nK)}\to s_{(nn)}K$ (denoted as $t_{b}$). For concreteness, the coupled-channel three-body equations for 
the {\it half-off-shell} amplitudes $t_{a}$ and $t_{b}$ are diagrammatically displayed (omitting the three-body 
contact terms for brevity) in Fig.~\ref{fig:diagram}~\cite{STM1,STM2,Canham:2008jd,Ando:2015fsa}.

It is straightforward to obtain these equations following the Feynman rules from the Lagrangian 
presented in the previous subsection. After $s$-wave projections, we obtain the following expressions:

\begin{widetext}
\begin{align}\hspace{-1cm}
    \quad t_{a}(p^{\prime},p;E) &=
    m_{K}\left\{\frac{1}{2p^{\prime}p} 
    \ln\left[\frac{p^{\prime 2}+p^{2}
    +ap^{\prime}p
    -2\mu_{(nK)}E
    }
    {p^{\prime 2}+p^{2}
    -ap^{\prime}p
    -2\mu_{(nK)}E
    }\right]-\frac{g_s(\Lambda)}{\Lambda^2}\right\} \nonumber \\*
    &\quad -\frac{m_{K}}{\pi\mu_{(nK)}}
    \int_{0}^{\Lambda} dl \,\,l^2
    \left\{\frac{1}{2p^{\prime}l} 
    \ln\left[\frac{p^{\prime 2}+l^{2}
    +ap^{\prime}l
    -2\mu_{(nK)}E}
    {p^{\prime 2}+l^{2}
    -ap^{\prime}l
    -2\mu_{(nK)}E}\right]-\frac{g_s(\Lambda)}{\Lambda^2}\right\}
    \frac{t_{a}(l,p;E)}{
    \frac{1}{a_{d(nK)}}
    -
    \sqrt{-2\mu_{nK}E
    +
    \frac{\mu_{(nK)}}{\mu_{n(nK)}}l^{2}
    -i0^{+}}
    } 
    \nonumber \\*
    &\quad -
    \frac{\sqrt{2}}{\pi}
    \int_{0}^{\Lambda} dl \frac{l}{p^{\prime}}  
    \ln\left[\frac{p^{\prime 2}+bl^{2}
    +p^{\prime}l
    -M_{n}E}
    {p^{\prime 2}+bl^{2}
    -p^{\prime}l
    -M_{n}E}\right]  
    \frac{t_{b}(l,p;E)}{
    \frac{1}{a_{s(nn)}}
    -
    \sqrt{-M_{n}E
    +
    \frac{M_{n}}{2\mu_{K(nn)}}l^{2}
    -i0^{+}}
    } 
\label{eq:ta}
\end{align}
and
\begin{align}
    \quad t_{b}(p^{\prime},p;E) &= 
    \frac{M_{n}}{\sqrt{2}p^{\prime}p}
    \ln
    \left[\frac{bp^{\prime 2}+p^{2}
    +p^{\prime}p-M_{n}E 
    }
    {bp^{\prime 2}+p^{2}
    -p^{\prime}p-M_{n}E 
    }\right] \nonumber \\
    &\quad 
    -\frac{M_{n}}{\sqrt{2}\pi \mu_{(nK)}}
    \int_{0}^{\Lambda} dl \frac{l}{p^{\prime}}
    \ln
    \left[\frac{bp^{\prime 2}+ l^{2}
    +p^{\prime}l-M_{n}E 
    }
    {bp^{\prime 2}+l^{2}
    -p^{\prime}l-M_{n}E 
    }\right]
    \frac{t_{a}(l,p;E)}{
    \frac{1}{a_{d(nK)}}
    -
    \sqrt{-2\mu_{(nK)}E+\frac{\mu_{(nK)}}{\mu_{n(nK)}}l^{2}
    -i0^{+}}
    } \,. 
\label{eq:tb}
\end{align}
\end{widetext}
Here $E$ is the total center-of-mass kinetic energy of the three-body system, $p$ and $p^{\prime}$ 
are the initial and final state momenta, and $\mu_{(nK)}(x)=M_{n}m_{K}(x)/\left[M_{n}+m_{K}(x)\right]$ 
is the meson-neutron reduced mass. The $s$-wave meson-neutron and neutron-neutron scattering lengths, 
$a_{d(nK)}\equiv a_{d(nK)}(x)$ and $a_{s(nn)}$, respectively, are given as
\begin{equation}
    \frac{1}{a_{d(nK)}(x)}
    = \frac{2\pi }{\mu_{(nK)}(x)\,g_{nK}}+\mu_{p.d.s.}\,,
\end{equation}
\begin{equation}
\frac{1}{a_{s(nn)}}
    = \frac{4\pi }{M_{n}g_{nn}}+\mu_{p.d.s.} ,
\end{equation}
when the power divergence subtraction scheme~\cite{Kaplan:1998tg,Kaplan:1998we,vanKolck:1998bw} is adopted 
with subtraction scale $\mu_{p.d.s.}\sim m_\pi$. Furthermore, we introduce the neutron-dihadron ($n+nK$)-reduced 
mass as $\mu_{n(nK)}(x)=M_{n}[M_{n}+m_{K}(x)]/\left[2M_{n}+m_{K}(x)\right]$, along with the two other mass dependent 
parameters: 
\begin{align}
    a(x)
    &= 
    \frac{2\mu_{(nK)}(x)}{m_{K}(x)}, \quad
    b(x)
    = 
    \frac{M_{n}}{2\mu_{(nK)}(x)}\,.
\end{align}
In the above expressions, we re-emphasized the dependence on the extrapolation parameter $x$ at a 
general unphysical point $x\neq 0,1$.
The coupled integral equations must be numerically solved to obtain the non-perturbative 
solutions to the three-body problem. It is important to note that these integral equations must be
regularized, e.g., by introducing the sharp momentum cut-off $\Lambda$ to remove ambiguities 
in the asymptotic phase with the cut-off taken to infinity~\cite{BHV98,DL63}. The resulting cut-off 
scale dependence is renormalized  by introducing the leading order three-body contact interaction 
term displayed in Eq.~\eqref{eq:L3body} with a cut-off dependent running coupling, 
$g_s(\mu=\Lambda)\propto\ln(\Lambda)$, leading to a RG limit cycle. 

\subsection{Asymptotic analysis}

\begin{figure}[tbp]
    \includegraphics[width=9.1cm, height=5.6cm]{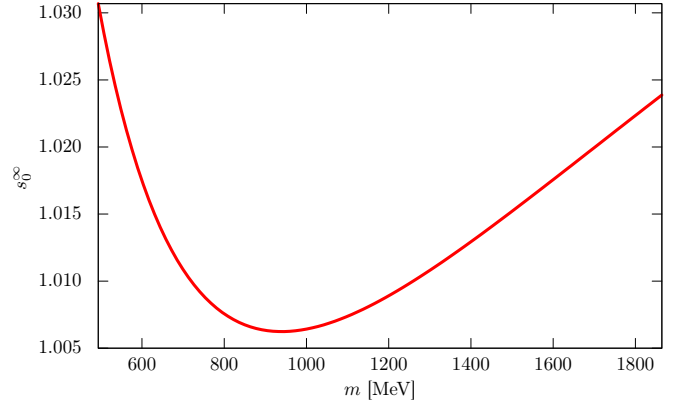}
    \caption{\label{fig:asymptotic}
    Flavored meson mass dependence on the asymptotic parameter $s_{0}^{\infty}$ for $m_{K^-}<m<m_{D^0}$. }
\end{figure}

In order to determine whether the three-body system exhibits the Efimov effect, it is instructive to examine the 
asymptotic behavior (both in the {\it scaling} as well as unitary limits) of the integral equations with the 
cut-off removed (i.e., $\Lambda\to \infty$) and the three-body contact interaction excluded (see
Refs.~\cite{Braaten:2004rn,AH04,Canham:2009zq} for details). 
In this limit, the scale of the off-shell momenta can be considered very large in comparison with 
the inverse two-body scattering lengths and the eigenenergies, $E=-B_d<0$ of the three-body bound 
state\footnote{
Here we use the notation, $B_d=-E>0$, to denote the {\it absolute} value of the trimer binding energy, measured with 
respect to three-particle break-up threshold. This is to distinguish $B_d$ from our notation for the {\it relative} 
trimer energy, $B_T=B_d-E_D>0$, measured from the particle-dimer break-up threshold energy of $E_D$ 
[also, see Eq.~\eqref{eq:ED}].}, i.e., $p^\prime,l \gg 1/a_{d(nK)},1/a_{s(nn)},\sqrt{2\mu_{(nK)}B_d},\sqrt{M_{n}B_{d}}$. In 
particular, the asymptotic solution to the amplitudes is expected to follow a power law: $t_{a,b}(p^\prime)\propto p^{\prime s-1}$. 
It is then straightforward to show that the coupled integral equations in this regime get reduced into a single 
transcendental equation which may be solved for the exponent $s$.
By using the meson masses in the physical limits, i.e., for the antikaon $K^{-}(x=0)$ and the charmed meson 
$D^{0}(x=1)$, as given in Table~\ref{tbl:hadronmass}, we obtain in each case a pair of imaginary solutions, 
$s=\pm is_{0}^{\infty}$, with the respective transcendental number $s^{\infty}_0$ given by 
\begin{align}
    s_{0}^{\infty}
    &= 
    \begin{cases}
    1.03069  & \text{for } K^{-}nn\,, \\
    1.02387  & \text{for }D^{0}nn\,. 
    \end{cases}
\end{align}
This confirms the previously obtained result by Braaten and Hammer (see Fig.\,52 of Ref.~\cite{Braaten:2004rn}).
In fact, the imaginary solutions of $s$ in the unphysical domain, i.e., $0< x< 1$, are {\it continuously connected} 
in the process of interpolating the flavored meson mass $m \equiv m_K(x)$ in between the physical limits $m_{K^-}(x=0)$ 
and $m_{D^0}(x=1)$, as depicted in Fig.~\ref{fig:asymptotic}. We thereby conclude that the coupled integral equations are 
ill-defined in the asymptotic limit which means that {\it formally} Efimov effect must be manifest in the three-body 
system, not only at the physical limits ($x=0,1$), but at {\it all} intermediate unphysical points. Of course one 
should bear in mind the formal nature of the above results which follow from a leading order asymptotic analysis 
corresponding to the scaling limit of all the two-body interactions. But the more robust physical realization of the 
Efimov spectrum, i.e., whether or not it lies within an {\it Efimov window}, evidently relies on the nature of possible 
crucial range effects. 

\subsection{Numerical results}

As already mentioned, the numerical evaluations of the three-body integral equation require the meson-neutron 
and neutron-neutron $s$-wave scattering lengths as the principal two-body input to our leading order EFT 
analysis. First, our numerical results are displayed with the spin-singlet $s$-wave $nn$ scattering length 
taken as its physical value, $a_{s(nn)}=-18.63$ fm~\cite{Chen2008}. Second, we try to demonstrate the remnant 
universal features of the non-asymptotic results in the physical sectors whereby we neglect the 
{\it absorptive} contributions of the decay/coupled channels for simplicity. For the purpose of 
extrapolating to the unphysical domain to probe the unitary limit, it is convenient to use our predicted 
scattering lengths in the ZCL, as displayed in Eq.~(\ref{eq:ZCL_ad}) in the physical limits ($x=0,1$), 
namely,
\begin{eqnarray}
a_{d(nK^{-})}&\equiv &a_{d(nK)}(x=0)=a_{0,K^{-}n}^{\rm ZCL}=-0.394\,\,\text{fm}\,,\nonumber \\
a_{d(nD^{0})}&\equiv &a_{d(nK)}(x=1)=a_{0,D^{0}n}^{\rm ZCL}=4.141\,\,\text{fm}.
\label{eq:ad_ZCL}
\end{eqnarray}
We thereby solve the non-asymptotic integral equations \eqref{eq:ta} and \eqref{eq:tb} to obtain the 
three-body energy eigenvalues as a function of the sharp cut-off regulator $\Lambda$.  

We first include the three-body force containing the coupling $g_s(\Lambda)$ to investigate into its approximate 
limit cycle nature at non-asymptotic momenta/energies. Later, we shall focus on the results excluding the 
three-body term to investigate into the behavior of the three-body eigenenergies of the physical sectors (i.e., 
$x=0,1$ limits), generated exclusively from the two-body dynamics. In particular, the role of the cut-off 
($\Lambda$) dependence of the results is investigated which can crucially determine whether the three-body 
bound state formation is formally supported in the low-energy EFT with zero-range approximation. Furthermore, 
in the context of the ZCL approach, a study of the first ``critical'' cut-off $\Lambda_c^{(n=0)}$ (defined 
later in the sub-section) excluding the three-body force is presented, whose variation with the 
extrapolation parameter $x$ reveals vital information on three-body universality. Finally, in association 
with our findings in the two-body sector, the above results may be used to predict the likelihood of realistic 
Efimov-like bound states. A brief description of our methodology in the numerical solution to the integral 
equations is presented in Appendix~\ref{sec:AppendixA}.

We then present our result for the RG limit cycle for the three-body contact interaction coupling $g_{s}(\Lambda)$ 
in the $K^-nn$ ($D^0nn$) system. For a proper estimation of this coupling a prior knowledge of a three-body observable, 
such as the $s$-wave three-particle (particle-dimer) scattering length or the trimer binding energy is required, 
neither of which is currently available. However, even in the absence of such a three-body datum, we may study 
the RG behavior of this coupling by fixing any presumably small/near-threshold value of the trimer binding energy, 
say, $B_d, B_T = 0.1$ MeV, and varying the cut-off scale $\Lambda$. Various qualitative features concerning three-body 
universality can be deduced thereof.

\begin{figure}[tbp]
    \centering
    \includegraphics[width=\columnwidth, height=5.5cm]{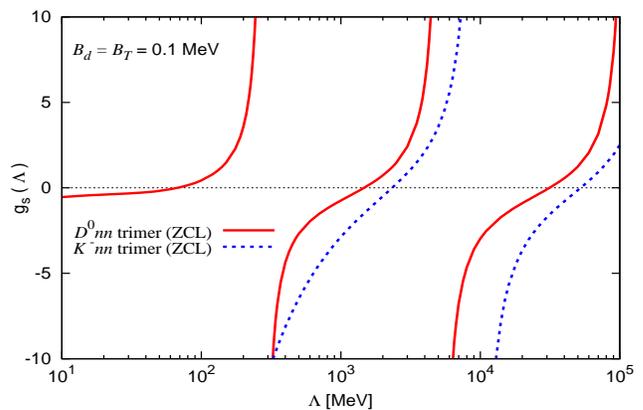}
\caption{\label{fig:g-lambda_zcl} 
RG Limit cycle for the three-body coupling $g_s$ as a function of the sharp momentum cut-off $\Lambda$. The pertinent 
integral equations are solved with the input two-body scattering lengths from Eq.~(\ref{eq:ad_ZCL}) in the zero coupling 
limit for fixed trimer eigenenergies, namely, $B_d=0.1$ MeV for $K^-nn$, measured with respect to the 
three-particle break-up threshold, and  $B_T=B_d-E_D= 0.1$ MeV for $D^0nn$, measured with respect to the particle-dimer 
break-up threshold energy of $E_D=1.82$ MeV. }
\end{figure}

Figure~\ref{fig:g-lambda_zcl} demonstrates the typical cyclic or quasi-log-periodic behavior of the coupling 
$g_s(\Lambda)$ for the $K^{-}nn$ and $D^0nn$ systems, reflecting an approximate RG limit cycle in each case. 
The periodic divergences of the coupling are associated with the sequence of successive formation of three-body 
bound states. For the $D^0nn$ system the first three bound states are evident for $\Lambda<10^5$ MeV, and for the 
$K^-nn$ system only the first two are visible in this range. The qualitative feature of the limit cycle plots remains 
unchanged whether the binding energy, $B_d$ ($B_T$), is chosen approaching the three-particle (particle-dimer) 
break-up threshold, as shown in the figure, or chosen somewhat away from the threshold; there is basically a slight 
downward and rightward shift of each curve with increasing $B_d$ ($B_T$) because the interaction becomes more and 
more attractive. 

For the running coupling $g_{s}(\Lambda)$, the characteristic periodicity, with $\Lambda = \Lambda^{(N)}$ 
$\forall N\in \mathbb{N}$,  can be expressed as
\begin{equation}
g_s(\Lambda^{(1)})
=g_s(\Lambda^{(N+1)}) 
\,\, {\rm with} \,\, 
\Lambda^{(N+1)}
=\Lambda^{(1)}\,{\rm exp}\left(\frac{N\pi}{s_0}\right)
\,\,, 
\end{equation} 
where $s_0$ is a real three-body  parameter reminiscent of the corresponding asymptotic limit cycle 
value of $s^\infty_0=1.03069$ ($s^\infty_0=1.02387$) that we found earlier for the $K^-nn$ ($D^0nn$) system. 
Physically, $s_0$ gives a measure of the residual approximate discrete scale invariance surviving in the 
non-asymptotic regime. In particular, the running coupling vanishes quasi-periodically at a discrete set of 
cut-off values, expressed as $\Lambda^{(N+1)}_0=\Lambda^{(1)}_0\,{\rm exp}\left(\frac{N\pi}{s_0}\right)$. 
It is to be noted that $s^\infty_0$ is a universal number, depending only on the gross features of the 
three-particle system, like ratios of the respective masses of the three particles involved, or the 
overall spin and isospin quantum numbers of the system. On the other hand, the non-universal number $s_0$ 
deviates from $s^\infty_0$ primarily due to cut-off dependent effects, implying an implicit dependence on 
$N$ itself, i.e., $s_0\equiv s^{(N)}_0$. Additional non-asymptotic parametric dependence on the trimer 
binding energy, three-body coupling and the two-body scattering lengths, can further influence the 
numerical value  of $s^{(N)}_0$. 
In other words, $s^{(N)}_0$ being the outcome of a numerical RG is purely numerical by nature, and 
thereby difficult to ascribe a unique definition in terms of a analytical expression. However, reasonably 
good estimates are obtained by taking the successive discrete cut-offs, say, $\Lambda^{(N)}_0$ and $\Lambda^{(N+1)}_0$ 
for vanishing coupling $g_s(\Lambda^{(N)}_0)=g_s(\Lambda^{(N+1)}_0)=0$, and using the closed-form definition: 
\begin{equation}
     s_0^{(N)}=\frac{\pi}{\ln\left(\frac{\Lambda^{(N+1)}_0}{\Lambda^{(N)}_0}\right)}\,. 
\end{equation} 
A crucial observation in this definition is that the sequence of the non-asymptotic numbers 
$s^{(N)}_0$, as obtained above, gradually converges to the asymptotic limit value $s^\infty_0$ 
beyond a suitably large $N>0$. Mathematically, stated:  
{\it for $\epsilon > 0$,\, $\exists\, N_0 \in \mathbb{N}$,\, such that}
\begin{equation}
\left| s_0^{(N)} - s_0^{\infty} \right| < \epsilon\,,\quad \forall \, N > N_0\,.
\end{equation}
An alternative method of extraction of the parameter $s_0$ is presented in Appendix~\ref{sec:AppendixB}.

\begin{figure}[tbp]
    \centering
    \includegraphics[width=\columnwidth, height=5.5cm]{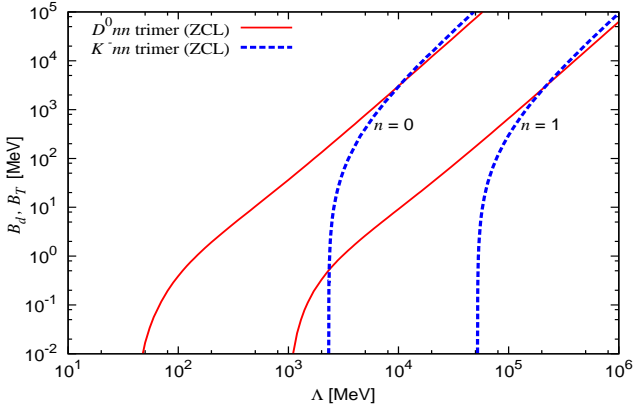}
\caption{\label{fig:bd-lambda_zcl} 
Binding energies for the ground ($n=0$) and first ($n=1$) excited level states as a function of the sharp 
momentum cut-off $\Lambda$, excluding the three-body force.  The pertinent integral equations are solved with the input 
two-body scattering lengths from Eq.~(\ref{eq:ad_ZCL}) in the zero coupling limit. For the $K^-nn$ system, the trimer 
binding energy $B_d$, is measured with respect to the three-particle break-up threshold, while for the $D^0nn$ system the 
binding energy, $B_T=B_d-E_D$, is measured with respect to the particle-dimer break-up threshold energy, $E_D=1.82$ MeV.}
\end{figure}

In the following numerical study, we shall exclude the three-body force since there is no 
three-body input datum available to fix the coupling $g_s$. In Fig.~\ref{fig:bd-lambda_zcl}, 
the results for the trimer energy of the physical systems, $K^-nn\, (x=0)$ and $D^0nn\, (x=1)$, 
are summarized. In the figure, the ground $(n=0)$ and the first excited $(n=1)$ state
binding energies, $-E=B_d>0$, of the $K^-nn$ {\it Borromean} trimer~\cite{Naidon:2016dpf} 
(with both $a_{d(nK^{-})},a_{s(nn)}<0$), measured with respect to the three-particle breakup threshold, 
are plotted as a function of $\Lambda$. The same figure also displays the corresponding 
eigenenergies, $B_T=B_d-E_D$, of the $D^0nn$ trimer state, measured with respect to the 
particle-dimer ($n+D^0n$) break-up threshold energy $E_D$. Note that the threshold value of the 
$D^0n$ dimer binding energy, which is given by
\begin{equation}
E_D=\frac{1}{2\mu_{(nD)}a^2_{d(nD)}},
\label{eq:ED}
\end{equation}
is obtained as $E=-E_D= -1.82$ MeV, using the $D^0n$ reduced mass $\mu_{(nD)}=\frac{M_nm_D}{M_N+m_D}$, and the 
corresponding ZCL scattering length given in Eq.~(\ref{eq:ad_ZCL}). It is to be noted that the value $-E_D$ 
closely matches with the value of the $D^0n$ bound state scattering amplitude pole position, $E_h=-1.85$ MeV 
(see Fig.~\ref{fig:trajectory_y1}). 

Figure~\ref{fig:bd-lambda_zcl} clearly indicates the increasing magnitude of the binding energies as the cut-off scale 
$\Lambda$ increases. 
In this figure, the first branch refers to the ground state of the $D^0nn$ ($K^-nn$) system, which appears 
as a shallow threshold state at the so-called ``critical'' cut-off value of $\Lambda^{\rm (n=0)}_{c}\simeq 38.4$ MeV 
($\simeq 2.3$ GeV). The state then becomes increasingly deeply bound with increasing cut-off, while a second branch (first 
excited state) appears at the critical value, $\Lambda^{\rm (n=1)}_c\simeq 1$ GeV ($\simeq 52.2$ GeV), which in its turn gets 
progressively deeper. Continuing in this manner an infinite tower of excited states emerges from the zero energy threshold 
as $\Lambda\to\infty$.\footnote{
Because here we exclude the three-body force, the critical point associated with the ground state, denoted here 
as $\Lambda_c^{(n=0)}$, corresponds to the first zero at $\Lambda=\Lambda_0^{(N=1)}$ of the three-body contact interaction, 
i.e., $g_s(\Lambda_0^{(N=1)})=0$ in Fig.~\ref{fig:g-lambda_zcl}, for fixed trimer binding energy.} Again, as pointed out 
earlier in the context of Fig.~\ref{fig:g-lambda_zcl}, only the ground state $D^0nn$ trimer is likely to satisfy conditions 
that may formally qualify it as an Efimov-like state.   


Let us finally study the flavor extrapolation from the $K^-nn$ system ($x=0$) to the $D^0nn$ system 
($x=1$). As seen in Sec.~\ref{subsec:extrapolation}, a continuous extrapolation of the meson-neutron 
system from $K^-n$ to $D^0n$ (see Fig.~\ref{fig:slengthreal}) in ZCL yielded an unitary limit 
of the meson-neutron interaction in the proximity of $x\sim 0.6$. Here excluding the three-body force we 
investigate the behavior of the ``critical point'' $\Lambda^{(0)}_c$, corresponding to the ground state ($n=0$), 
as a function of the extrapolation parameter $x$ for fixed (absolute or relative) trimer binding energy 
($B_d$ or $B_T$). In other words, we perform an extrapolation in the interval $x \in [0,0.615)$, i.e., along 
the three-particle break-up threshold with a fixed value of $B_d$, and continue further in the interval 
$x \in (0.615,1]$, i.e., along the particle-dimer break-up threshold with a fixed value of $B_T=B_d-E_D$, 
where $E_D\equiv E_D(x)$ in the latter interval is itself $x$ dependent. It may be noted that level energies 
chosen too close to the thresholds can lead to numerical instabilities in the vicinity of the unitary limit. 
We have, therefore, preferred non-zero but near threshold binding energies, say, $B_d,B_T\gtrsim 0.001$ MeV 
to suit our purpose of demonstration. 

\begin{figure}[tbp]
    \centering
    \includegraphics[width=\columnwidth, height=5.5cm]{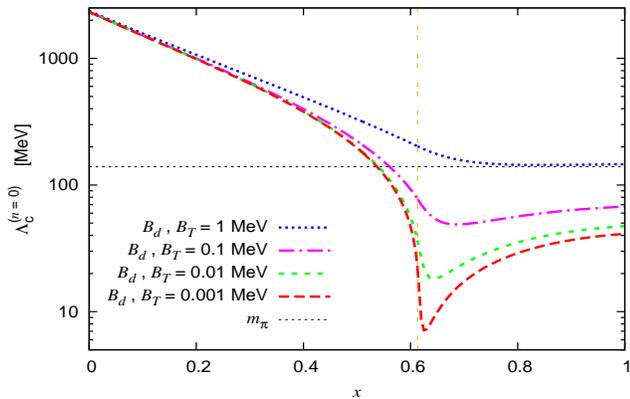}
\caption{\label{fig:lambdac-x} 
Extrapolation of the critical cut-off $\Lambda^{(0)}_c$, associated with the ground state 
($n=0$) for fixed binding energies ($B_d$ or $B_T$) of the meson-neutron-neutron bound state, 
measured along the three-particle or particle-dimer break-up threshold. The double 
dashed vertical line at $x = 0.615$ denotes the unitary limit. The plot corresponds to  
the zero coupling limit with input scattering lengths, as given in Eq.~(\ref{eq:ad_ZCL}).
The three-body force is excluded in these results.}
\end{figure}

Figure~\ref{fig:lambdac-x} displays our results where the meson mass and the $s$-wave meson-neutron scattering 
length are simultaneously extrapolated from $x=0$ to $x=1$.
The fact that there is a large change ($\sim 3$ orders of magnitude) in the critical cut-off along $x \in [0,0.615)$ 
compared to a rather nominal change along $x \in (0.615,1]$, suggests that the $D^0nn$ system is much more likely 
to be found in the domain of three-body universality and yields an Efimov-like bound state in comparison to 
the $K^-nn$ system. This reflects the much larger scattering length, $a_{0,D^0n}^{\rm ZCL}\sim 4$ fm, than that in the 
strangeness sector, $a_{0,K^-n}^{\rm ZCL}\sim -0.4$ fm. The horizontal dotted line in the above figure indicates the 
{\it one-pion threshold}, the upper cut-off of the pionless EFT employed in this analysis.\footnote{
In fact, it may be noted that the cut-off scale of the short-distance two-body interaction should be taken 
still larger since the {\it one-pion exchange} process is forbidden in the $K^{-}n$ or $D^{0}n$ system.} A direct 
comparison with the behavior of the critical cut-off indicates that a $D^0nn$ ground state trimer is clearly supported 
in the realm of the low-energy EFT framework, lying well below the hard scale ${}^{\pi\!\!\!/}\Lambda \sim m_\pi$, while 
the same for the $K^-nn$ system lies far above the applicability of the low-energy EFT, and thus, can be considered 
effectively unbound by Efimov attraction.

\section{Discussion and summary}\label{sec:summary}

In this work, we have explored the two and three-body universal physics associated with a 
three-particle cluster state of two neutrons and a flavored meson, i.e., an antikaon 
$K^{-}$ or a $D^{0}$ meson. We have demonstrated that the meson-neutron scattering length 
can become infinitely large in an idealized limit, around which universality governs the 
physics of two-body and three-body systems. This offers an interesting prospect that a possible 
remnant of this universal physics may be observed in the physical hadronic systems.

First, we have studied the two-body meson-neutron systems. Motivated by the experimental 
evidences that the $K^{-}n$ ($D^{0}n$) system has no quasi-bound state (one quasi-bound state), 
we construct a theoretical coupled-channel model to describe both $K^{-}n$ and $D^{0}n$ systems 
with a single flavor extrapolation parameter $x$. This model enables the extrapolation to 
unphysical quark masses between the strange $(x=0)$ and the charm $(x=1)$ flavors limits, providing 
a natural mechanism to tune the strength of the meson-neutron interaction. While the universal 
features are not very prominent due to the coupled-channel effects, it was shown that the 
meson-neutron interaction can reach resonant conditions with large two-body scattering length 
around $x\sim 0.6$ in the zero coupling limit (ZCL) with the channel couplings switched off.
We further investigated two-body universality by comparing the eigenmomentum of the system 
with that predicted solely by the two-body scattering lengths. 
This led to the identification of the entire region $x\gtrsim 0.3$ as the two-body universal 
window in the ZCL, 
including the charm sector ($x=1$). Moreover, the universal prediction reasonably works in the 
full coupled-channel model with complex meson-neutron scattering lengths in the extrapolation region 
above $x\gtrsim 0.8$. This indicates that universality governs not only the idealized ZCL scenario 
but also the physical $D^0n$ system, both being continuously connected. One can then conclude that 
the physical charm sector is likely to follow the predictions of universality, even with the couplings 
to the decay channels included.

Next, we investigated three-body universality by employing a low-energy cluster EFT.
In a simplified approach, we especially neglected the absorptive influence of sub-threshold decay channels 
which amounts to considering the $s$-wave meson-neutron scattering lengths to be real-valued. A leading order 
EFT analysis allowed us to investigate the meson-neutron-neutron system in the scaling limit (zero-range 
approximation) where the universal physics was determined only by the two $s$-wave scattering lengths $a_{d(nK)}$ 
and $a_{s(nn)}$. In this work $a_{s(nn)}$ was kept fixed to the physical value, while $a_{d(nK)} \equiv a_{d(nK)}(x)$ 
was varied as a function of $x$. 
The introduction of a sharp momentum cut-off $\Lambda$ in the integral equations led to the breaking of 
continuous scale invariance resulting in the introduction of range-like effects. In this context, it may be 
noted that effective range corrections were not explicitly considered in this qualitative leading order analysis. 
We anticipate that the range correction would not be significant in the charm sector based on the two-body analysis 
in Figs.~\ref{fig:eigenmomentumcomp} and \ref{fig:eigenmomentumreal}. For definite conclusion, however, it is important 
to explicitly examine the range correction, which is left as a future prospect of this work.
 
An immediate consequence of the breaking of continuous scale invariance in the three-body sector is the 
appearance of a discrete scaling symmetry in the solutions to the integral equations that can be associated 
with an asymptotic RG limit cycle. In this context, Fig.~\ref{fig:asymptotic} suggests that the asymptotic limit 
cycle between the $K^-nn$ and $D^0nn$ systems continuously exists for all intermediate values of $x$ so that the 
respective asymptotic parameters $s^\infty_0=1.03069$ (strange sector) and $s^\infty_0=1.02387$ (charm sector) are 
smoothly connected. This leads us to conclude that by neglecting the influence of decay channels and range corrections, 
the Efimov effect can always be manifested in the solutions to the three-body coupled integral equations. In fact, 
from Fig.~\ref{fig:g-lambda_zcl}, we confirmed approximate RG limit cycle behavior (with quasi-log-periodicity) 
of the three-body running coupling $g_s(\Lambda)$ in the non-asymptotic solutions for each system. 

Next, the binding energies of the three-body systems in the physical limits were obtained as a function of the cut-off 
$\Lambda$. For the sake of simplicity we excluded the dependence on the three-body force with the unknown coupling 
$g_s$. We found that the binding energies increase with increasing $\Lambda$, and the various level states emerge in 
order from the zero energy threshold, starting from a deepest (ground) state that appears at a certain critical value, 
$\Lambda=\Lambda^{(0)}_c$. It may be noted that a larger value of the momentum cut-off leads to additional inclusion of 
ultraviolet physics into the theory arising from high energy modes. This evidently means greater attraction in the system 
leading to larger three-body binding energies and the emergence of new level states. It clearly emerges from our results 
that the Efimov spectrum is much steeper on the negative side of the scattering lengths. With the much smaller critical 
cut-offs for the $D^0nn$ level states than those of the $K^-nn$ system, it is straightforward to conclude that the $D^0nn$ 
trimer states are manifested more easily. 

Finally, as shown in Fig.~\ref{fig:lambdac-x}, the behavior of the lowest critical cut-off $\Lambda_c^{(0)}$ 
associated with the emergence of near threshold (three-body) ground states was obtained as a function of $x$ 
using the extrapolated meson-neutron scattering length in the ZCL. Clearly with smaller and smaller chosen 
values of $B_d$ or $B_T$, $\Lambda_c^{(0)}$ converged to the point 
($x\simeq 0.6$,\,$\Lambda_c^{(0)}\simeq 0$), at which near threshold states were realized corresponding 
to the vanishing scale of two-body unitarity. As $x$ is taken away from unitarity toward the physical limits, 
$x=0,1$, $\Lambda_c^{(0)}$ becomes larger, indicating larger two-body interaction strengths needed to form 
three-body bound states. Clearly, the $x=1$ physical limit ($D^0nn$ system) is located much closer to the 
unitary limit at $x\simeq 0.6$ than the $x=0$ physical limit ($K^-nn$ system). This naturally indicates 
that a $D^0nn$ ground state trimer can be realized in the ZCL in pionless EFT framework with $\Lambda_c^{(0)}$ 
lying well below the pion mass. In fact, this is a straightforward consequence of the large $D^0n$ scattering length 
in the ZCL, i.e., $a_{0,D^0n}^{\rm ZCL}\sim 4$ fm, which traces back to the existence of $\Sigma_c(2800)$ near the $D^0n$ 
threshold. While for the $K^-nn$ system, with $\Lambda_c^{(0)}\gtrsim 2$ GeV, no physically realizable mechanism in the 
context of a low-energy EFT can generate sufficient interaction strength to form bound states. In a sense, the much 
steeper Efimov spectrum for the $K^-nn$ system is consistent with the fact that such Borromean Efimov trimers are by 
nature extremely difficult to form.     

In retrospect, through a simplistic qualitative analysis we have presented here a rather idealistic scenario 
whereby universal Efimov physics may be realized in a meson-neutron-neutron system, despite the smallness 
of the meson-neutron scattering lengths. Such a scenario may very well be unrealistic given the rather delicate 
nature of the Efimov-like physics which may easily become obscured or wiped out altogether due to range and 
coupled-channel effects with $C_{i1}=C_{1i}\neq 0$ ($i=2,3,4$) restored. Nevertheless, it may still be somewhat 
interesting to pursue further studies to assess the exact nature of the $D^{0}nn$ system which has never been 
previously considered as a possible candidate for a hadronic molecule. Having said that, it must also be borne 
in mind that the dynamics of the sub-threshold decay channels which were systematically neglected in this work 
can play a vital role to ultimately decide whether the physical $D^0nn$ system supports a quasi-bound state.  
With sizable imaginary parts in the meson-neutron scattering lengths, the dynamical effect of the decay channel 
should be carefully checked in a more robust coupled channel framework. This may be useful to gain better insights 
into the universal aspects of three-body dynamics, especially in the presence of two-body quasi-bound subsystems 
with large couplings to open channels. Thus, in regard to drawing definitive conclusions, the present leading order 
EFT analysis being too simplistic should either be further elaborated by including decay channels (employing complex 
scattering lengths) followed by sub-leading order effective range corrections, or by employing rigorous few-body 
techniques combined with realistic two-body potentials, e.g., as pursued in Ref.~\cite{Bayar:2012dd}. 


\appendix

\section{Solution via numerical integration}\label{sec:AppendixA}

We present here some details of our methodology used in this work to numerically solve the integral 
equations \eqref{eq:ta} and \eqref{eq:tb} (excluding the three-body force) in order to obtain the 
eigenenergies, $-E=B_d>0$, as a function of the sharp momentum cut-off $\Lambda$. For this purpose, 
we are only required to solve the homogeneous version of the coupled integral equations written in 
the general linearized form 
\begin{eqnarray}
t_a&=&A\otimes t_a+B\otimes t_b\,,\nonumber\\
t_b&=&C\otimes t_a+D\otimes t_b\,,
\end{eqnarray}
or equivalently,
\begin{eqnarray}
[x]={\bf X}\otimes [x]
\end{eqnarray}
with $[x]$, the column eigenvector and ${\bf X}$, the transformation matrix, defined as
\begin{eqnarray} 
[x]=
  \begin{bmatrix}
t_a\\ 
t_b
\end{bmatrix}\,,\quad
{\bf X}=
  \begin{bmatrix}
    A & B \\
    C & D 
  \end{bmatrix}\,.
\end{eqnarray}
In the present problem the elements of the $2\times 2$ transformation matrix 
${\bf X}(p^{\prime},l)$ are given by
\begin{eqnarray}
A(p^{\prime},l)&=& \frac{m_K}{\pi \mu_{(nK)}}\frac{l^2\,K_{(K)}(p^{\prime},l;E)}{\bigg[-\frac{1}{a_{d(nK)}}
+\sqrt{\frac{\mu_{(nK)}}{\mu_{n(nK)}}l^2-2\mu_{(nK)}E}\bigg]}\,,\nonumber\\
B(p^{\prime},l)&=& \frac{2\sqrt{2}}{\pi}\frac{l^2\,K_{(n1)}(p^{\prime},l;E)}{\bigg[-\frac{1}{a_{s(nn)}}
+\sqrt{\frac{M_n}{2\mu_{K(nn)}}l^2-M_nE}\bigg]}\,,\nonumber\\
C(p^{\prime},l)&=& \frac{\sqrt{2}M_n}{\pi \mu_{(nK)}}\frac{l^2\,K_{(n2)}(p^{\prime},l;E)}{\bigg[-\frac{1}{a_{d(nK)}}
+\sqrt{\frac{\mu_{(nK)}}{\mu_{n(nK)}}l^2-2\mu_{(nK)}E}\bigg]}\,,\nonumber\\
D(p^{\prime},l)&=& 0\,\,,
\end{eqnarray}  
where the $s$-wave projected one-meson exchange propagator is given as
\begin{equation}
K_{(K)}(p^{\prime},l;E)=\frac{1}{2lp^{\prime}}\ln\left[\frac{p^{\prime 2} + l^2 + ap^{\prime}l
- 2\mu_{(nK)}E}{p^{\prime 2} + l^2 - ap^{\prime}l - 2\mu_{(nK)}E}\right],
\end{equation}
while the one-neutron exchange interaction for the elastic and inelastic 
channels are parametrized by the two propagators:
\begin{eqnarray}
K_{(n1)}(p^{\prime},l;E)&=&\frac{1}{2lp^{\prime}}\ln\left[\frac{p^{\prime 2} + bl^2 + p^{\prime}l
- M_{n}E}{p^{\prime 2} + bl^2 - p^{\prime}l - M_{n}E}\right],\nonumber\\
K_{(n2)}(p^{\prime},l;E)&=&\frac{1}{2lp^{\prime}}\ln\left[\frac{bp^{\prime 2} + l^2 + p^{\prime}l
- M_{n}E}{bp^{\prime 2} + l^2 - p^{\prime}l - M_{n}E}\right].\nonumber\\
\end{eqnarray}
The integrations are evaluated employing the method of Gaussian quadrature where the integral over the 
loop momentum variable $l$ is replaced by a summation over $N$ Gauss-Legendre weights $w_j$, and correspondingly 
choosing $N\times N$ intermediate mesh points $(p^{\prime}_i,l_j)$ between the integration limits $[0,\Lambda]$, i.e.,
\begin{eqnarray}
\otimes\equiv \int^{\Lambda}_0dl\to \sum^{N}_{j=0}w_j\,.
\end{eqnarray}
To achieve good numerical convergence with a relatively small number of quadrature points, 
say, $N=96$, we find it convenient to reparametrize the integration variable in the following way:
First we assume that the momenta $l,\Lambda$ are re-scaled to dimensionless numbers by removing 
the dimensional parts, i.e., $l=\tilde{l}[\mu]$, $\Lambda=\tilde{\Lambda}[\mu]$, where $[\mu]$ 
has unit magnitude carrying the dimension of momentum, i.e., mass dimension 1, while $\tilde{l}, \tilde{\Lambda}$ 
are dimensionless numbers. Now, we may change the integration variable from $\tilde{l}\to\zeta$ as
\begin{eqnarray}
\tilde{l}=\frac{\zeta}{(1-\zeta)^7}\,, \quad d\tilde{l}=d\zeta\frac{1+6\zeta}{(1-\zeta)^8}\,,
\end{eqnarray}
so that the integral equations are reduced to a coupled system of $2N$ linear algebraic 
equations for $i=1,\cdots,N$:  
\begin{eqnarray}
\begin{bmatrix}
t_a(p^{\prime}_i,p;E)\\ 
t_b(p^{\prime}_i,P;E)
\end{bmatrix}
&=&\sum^N_{j=1}\tilde{w}_j\frac{1+6\zeta_j}{(1-\zeta_j)^8}\nonumber\\
&&\times\begin{bmatrix}
    A(p^{\prime}_i,\zeta_j) & B(p^{\prime}_i,\zeta_j) \\
    C(p^{\prime}_i,\zeta_j) & D(p^{\prime}_i,\zeta_j) 
  \end{bmatrix}
\begin{bmatrix}
t_a(\zeta_j,p;E)\\ 
t_b(\zeta_j,p;E)
\end{bmatrix}\,,\nonumber\\
\end{eqnarray}
for a given initial center-of-mass kinetic energy $E$. The new Gaussian weights $\tilde{w}_j$ 
corresponding to $N$ intermediate points $\zeta_j$ between the new integration limits $[0,\lambda]$ 
are related to the previously defined ones by
\begin{eqnarray}
w_j=\tilde{w}_j\frac{1+6\zeta_j}{(1-\zeta_j)^8}\,,\quad \tilde{\Lambda}=\frac{\lambda}{(1-\lambda)^7} \,.
\end{eqnarray}
On integration, we restore the original form of the momentum cut-off from $\lambda \to \Lambda$. 
In this way we may expand the domain of numerical integration to include momenta up to 12 orders 
of magnitude, maintaining sufficient numerical accuracy.

\begin{figure}[tbp]
    \centering
    \includegraphics[width=\columnwidth, height=5.9cm]{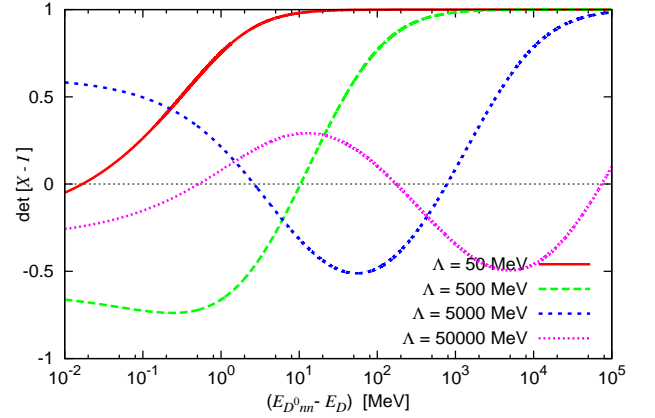}
\caption{\label{fig:det-lambda} 
${\rm Det} [X-I]$ vs energy, $-E=(E_{D^0nn}-E_D)$, plotted for the $D^0nn$ system for different sharp momentum 
cut-offs $\Lambda$. $E_D= 1.82$ MeV is the threshold energy for break-up into a $D^0n$ dimer and a neutron,
obtained using the $D^0nn$ scattering length in Eq.~(\ref{eq:ad_ZCL}), as the input. The nodes (zeros) correspond 
to eigenenergies of the trimer level states. Here solutions up to the first three bound states are displayed.}
\end{figure}
\begin{figure*}[tbp]
    \centering
    \includegraphics[width=18cm, height=5.7cm]{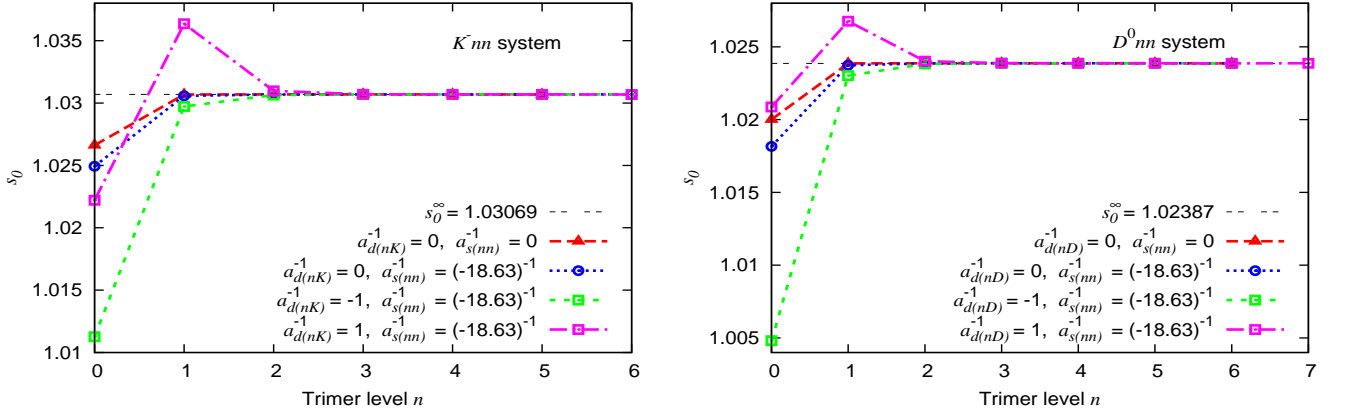}
\caption{\label{fig:convergence} 
Convergence of the three-body parameter $s_0$ for a fixed eigenenergy, $B_d=B_T=0.1$ MeV, for the $K^-nn$ 
and $D^0nn$ trimers. The points corresponding to equal values of the two-body (inverse) scattering lengths 
(in fm$^{-1}$) are connected by straight lines in order to guide the eyes. The long-dashed (red) lines that 
correspond to true two-body unitarity shows the quickest asymptotic convergence. The other lines correspond 
to varying the meson-neutron scattering lengths with the neutron-neutron scattering length fixed at the physical 
value, $a_{s(nn)}=-18.63$ fm. To achieve reasonable convergence up to the $n=7{\rm th}$ excited state, the number 
of quadrature points used is $N=300$.  }
\end{figure*}

The consistency of the above system of linear equations imply ${\rm det}(X-I)=0$, the solution (zeros) to which 
determines the eigenenergies for the two-body inputs, namely, the $s$-wave meson-neutron and neutron-neutron scattering 
lengths. As a demonstration, Fig.~\ref{fig:det-lambda} displays the plot of ${\rm det} [X-I]$ vs energy 
$-E=(E_{D^0nn}-E_D)>0$ for the $D^0nn$ system for different sharp cut-offs ($\Lambda$), using the $D^0n$ scattering 
length in the ZCL [see Eq.~(\ref{eq:ad_ZCL})].\footnote{
It is also clearly revealed in Fig.~\ref{fig:bd-lambda_zcl}, that trimer eigenenergies $B_T$ exist well below  
$\Lambda \lesssim 100$ MeV for the $D^0nn$ system.} The figure depicts the kinematic sector with $E_{D^0nn}$ above the 
particle-dimer break-up threshold energy, $E_{D}\simeq 1.82$~MeV, as given by Eq.~(\ref{eq:ED}).
Below this threshold, i.e., $E_{D^0nn}<E_D$, the three-body system energetically favors dissociating into a 
$D^0n$ dimer and a neutron. 
The trimer level states emerge above this threshold which correspond to the zeros of the 
individual oscillatory curves. 

\section{Convergence test of numerical integration}\label{sec:AppendixB}

As a simple qualitative test of the asymptotic convergence of the limit cycle behavior, due to both
cut-off dependent artifacts, as well as instabilities arising from the numerical integration, it may 
be instructive to investigate the behavior of the three-body parameter $s_0 \equiv s_0^{(n)}$ in the non-asymptotic 
solution. Note the implicit dependence of this parameter on the $n{\rm th}$ critical cut-off, indicated by the  
superscript $n$. Referring back to Fig.~\ref{fig:bd-lambda_zcl} in the main text, if one continues to 
very large values of the cut-off $\Lambda$, a discrete sequence of critical points is obtained following 
a rough logarithmic periodicity $\propto \pi/s_0^{(n)}$. However, the crucial observation is that as $\Lambda\to \infty$ 
this ratio must eventually converge to the {\it universal} value $\pi/s^\infty_0$, where $s^\infty_0$ is a transcendental 
number characterizing the asymptotic limit cycle, independent of the scattering lengths and energy eigenvalues of the 
bound state. In other words, if $\Lambda^{(n)}_c$ and $\Lambda^{(n+1)}_c$ are successive critical points corresponding to 
the $n{\rm th}$ and $(n+1){\rm th}$ level states ($n=0$ being the ground state) 
emerging at threshold then,
\begin{equation}
     s_0^{(n)}=\frac{\pi}{\ln\left(\frac{\Lambda^{(n+1)}_c}{\Lambda^{(n)}_c}\right)} , \quad \lim_{n\to\infty} s_0^{(n)} = s^\infty_0\,.
\label{eq:s0}
\end{equation} 
This behavior is plotted in Fig.~\ref{fig:convergence} for the case of a possible $K^-nn$ bound Borromean 
state (left plot) and a $D^0nn$ bound state (right plot), with fixed energy eigenvalue chosen sufficiently 
close to the respective thresholds. For the first few values of $n$, the ratios of the critical cut-offs 
display significant deviations due to finite cut-off effects. In fact, it is to be noted that the ground ($n=0$) 
state of three-body systems is predominantly known to display maximal deviations from universal character of 
the Efimov spectrum, lying far beyond the universal window around the unitary limit. However, both plots 
suggest a rapid convergence of $s_0$ towards the corresponding asymptotic values, $s^\infty_0=1.03069$ and $1.02387$, 
for $n\geq 2$. This is evident in all cases where different positive and negative two-body (inverse) scattering 
lengths are chosen approaching the unitary limit from either direction, with the meson masses kept fixed to the 
respective physical values. In particular, the long-dashed (red) lines in the figure that correspond to true 
unitarity of the two-body sub-systems show the quickest asymptotic convergence. But despite this apparent 
convergence, beyond a certain number $n$ the numerical precision of the integration routine becomes poor, 
resulting in deviations again from the asymptotic values. In that case, numerical accuracy is systematically 
improved by an appropriate choice of a large value of the number of quadrature points $N$, e.g., $N\geq 300$ 
may be necessary to obtain reasonable convergence above the $n=7{\rm th}$ excited Efimov state in the above figure, 
where the corresponding critical cut-off exceeds $\Lambda^{(n>7)}_c\gtrsim 10^{13}$ MeV. But the practical 
disadvantage of choosing a large $N$ is the  dramatic increase in the computational time which is undesirable. 
However, in the context of our low-energy EFT analysis the numerical integration is naturally restricted to 
low momentum cut-off values where an optimal choice, e.g., $N=96$, yields reasonably good convergence up to 
the $n=5{\rm th}$ excited state through the additional reparametrization of the integration variable, as described 
in Appendix~\ref{sec:AppendixA}.\\

\section*{Acknowledgments}
The authors are grateful to W. Weise, U. van Kolck, and G. Meher 
for useful discussions.
U.R. gratefully acknowledges the hospitality of YITP during 
the initial stages of the work. 
S.I.A. acknowledges the hospitality of IIT Guwahati during 
the final stage of the work.
The work of Y.K. was supported in part 
by JSPS KAKENHI Grant No. JP17J04333.
The work of S.I.A. was supported 
by the Basic Science Research Program through the National Research
Foundation of Korea funded by the Ministry of Education of Korea
(Grant No. NRF-2016R1D1A1B03930122).
The work of T.H. was supported in part 
by JSPS KAKENHI Grant No. JP16K17694 
and by the Yukawa International Program for Quark-Hadron Sciences (YIPQS).    


%

\begin{thebibliography}{10}

\bibitem{Hyodo:2011ur}
T.~Hyodo and D.~Jido,
\newblock Prog. Part. Nucl. Phys. {\bf 67}, 55 (2012).

\bibitem{Gal:2016boi}
A.~Gal, E.~V.~Hungerford and D.~J.~Millener,
\newblock Rev. Mod. Phys. {\bf 88}, 035004 (2016).

\bibitem{PL7.288}
Y.~Nogami,
\newblock Phys. Lett. {\bf 7}, 288 (1963).

\bibitem{Akaishi:2002bg}
Y.~Akaishi and T.~Yamazaki,
\newblock Phys. Rev. C {\bf 65}, 044005 (2002).

\bibitem{Shevchenko:2006xy}
N.~V.~Shevchenko, A.~Gal and J.~Mares,
\newblock Phys. Rev. Lett. {\bf 98}, 082301 (2007).

\bibitem{Shevchenko:2007ke}
N.~V.~Shevchenko, A.~Gal, J.~Mares and J.~Revai,
\newblock Phys. Rev. C {\bf 76}, 044004 (2007),.

\bibitem{Ikeda:2007nz}
Y.~Ikeda and T.~Sato,
\newblock Phys. Rev. C {\bf 76}, 035203 (2007).

\bibitem{Ikeda:2008ub}
Y.~Ikeda and T.~Sato,
\newblock Phys. Rev. C {\bf 79}, 035201 (2009).

\bibitem{Yamazaki:2007cs}
T.~Yamazaki and Y.~Akaishi,
\newblock Phys. Rev. C {\bf 76}, 045201 (2007).

\bibitem{Dote:2008in}
A.~Dote, T.~Hyodo and W.~Weise,
\newblock Nucl. Phys. A {\bf 804}, 197 (2008).

\bibitem{Dote:2008hw}
A.~Dote, T.~Hyodo and W.~Weise,
\newblock Phys. Rev. C {\bf 79}, 014003 (2009).

\bibitem{Wycech:2008wf}
S.~Wycech and A.~Green,
\newblock Phys. Rev. C {\bf 79}, 014001 (2009).

\bibitem{Ikeda:2010tk}
Y.~Ikeda, H.~Kamano and T.~Sato,
\newblock Prog. Theor. Phys. {\bf 124}, 533 (2010).

\bibitem{Barnea:2012qa}
N.~Barnea, A.~Gal and E.~Liverts,
\newblock Phys. Lett. B {\bf 712}, 132 (2012).

\bibitem{Ohnishi:2017uni}
S.~Ohnishi, W.~Horiuchi, T.~Hoshino, K.~Miyahara and T.~Hyodo,
\newblock Phys.\ Rev.\ C {\bf 95}, 065202 (2017).

\bibitem{Revai:2016muw} 
J.~Revai,
Phys.\ Rev.\ C {\bf 94},  054001 (2016)
  
\bibitem{Hoshino:2017mty} 
T.~Hoshino, S.~Ohnishi, W.~Horiuchi, T.~Hyodo and W.~Weise,
\newblock Phys.\ Rev.\ C {\bf 96}, 045204 (2017).

\bibitem{Uchino:2011jt}
T.~Uchino, T.~Hyodo and M.~Oka,
\newblock Nucl. Phys. A {\bf 868-869}, 53 (2011).

\bibitem{Hosaka:2016ypm}
A.~Hosaka, T.~Hyodo, K.~Sudoh, Y.~Yamaguchi and S.~Yasui,
\newblock Prog. Part. Nucl. Phys. {\bf 96}, 88 (2017).

\bibitem{Brambilla:2010cs}
N.~Brambilla {\em et~al.},
\newblock Eur. Phys. J. C {\bf 71}, 1534 (2011).

\bibitem{Hosaka:2016pey}
A.~Hosaka, T.~Iijima, K.~Miyabayashi, Y.~Sakai and S.~Yasui,
\newblock PTEP {\bf 2016}, 062C01 (2016).

\bibitem{Olive:2016xmw}
C.~Patrignani,
\newblock Chin. Phys. {\bf C40}, 100001 (2016).

\bibitem{Hofmann:2005sw}
J.~Hofmann and M.~F.~M.~Lutz,
\newblock Nucl. Phys. A {\bf 763}, 90 (2005).

\bibitem{Mizutani:2006vq}
T.~Mizutani and A.~Ramos,
\newblock Phys. Rev. C {\bf 74}, 065201 (2006).

\bibitem{GarciaRecio:2008dp}
C.~Garcia-Recio {\em et~al.},
\newblock Phys. Rev. D {\bf 79}, 054004 (2009).

\bibitem{Haidenbauer:2010ch}
J.~Haidenbauer, G.~Krein, U.-G.~Meissner and L.~Tolos,
\newblock Eur. Phys. J. A {\bf 47}, 18 (2011).

\bibitem{Yamaguchi:2011xb}
Y.~Yamaguchi, S.~Ohkoda, S.~Yasui and A.~Hosaka,
\newblock Phys. Rev. D {\bf 84}, 014032 (2011).

\bibitem{Yamaguchi:2013ty}
Y.~Yamaguchi, S.~Ohkoda, S.~Yasui and A.~Hosaka,
\newblock Phys. Rev. D {\bf 87}, 074019 (2013).

\bibitem{Bayar:2012dd}
M.~Bayar {\em et~al.},
\newblock Phys. Rev. C {\bf 86}, 044004 (2012).

\bibitem{Yamaguchi:2013hsa}
Y.~Yamaguchi, S.~Yasui and A.~Hosaka,
\newblock Nucl. Phys. A {\bf 927}, 110 (2014).

\bibitem{Braaten:2004rn}
E.~Braaten and H.-W.~Hammer,
\newblock Phys. Rept. {\bf 428}, 259 (2006).

\bibitem{Naidon:2016dpf}
P.~Naidon and S.~Endo,
\newblock Rept. Prog. Phys. {\bf 80}, 056001 (2017).

\bibitem{Efimov:1970zz}
V.~Efimov,
\newblock Phys. Lett. B {\bf 33}, 563 (1970).

\bibitem{Braaten:2003eu}
E.~Braaten and H.~Hammer,
\newblock Phys. Rev. Lett. {\bf 91}, 102002 (2003).

\bibitem{Konig:2016utl} 
S.~K\"onig, H.~W.~Grie\ss hammer, H.~W.~Hammer and U.~van Kolck,
Phys.\ Rev.\ Lett.\  {\bf 118}, no. 20, 202501 (2017).

\bibitem{Canham:2009zq}
D.~L.~Canham, H.~W.~Hammer, and R.~Springer, 
\newblock Phys. Rev. D {\bf 80}, 014009 (2014).
  
\bibitem{Braaten:2003he}
E.~Braaten and M.~Kusunoki,
\newblock Phys. Rev. D {\bf 69}, 074005 (2004).

\bibitem{Hyodo:2013zxa}
T.~Hyodo, T.~Hatsuda and Y.~Nishida,
\newblock Phys. Rev. C {\bf 89}, 032201 (2014).

\bibitem{Ando:2015fsa}
S.-I.~Ando, U.~Raha and Y.~Oh,
\newblock Phys. Rev. C {\bf 92}, 024325 (2015).

\bibitem{Barnea:2013uqa} 
N.~Barnea, L.~Contessi, D.~Gazit, F.~Pederiva and U.~van Kolck,
Phys.\ Rev.\ Lett.\  {\bf 114}, no. 5, 052501 (2015).
  
\bibitem{Bazzi:2011zj}
 M.~Bazzi {\em et~al.} (SIDDHARTA Collaboration),
\newblock Phys. Lett. B {\bf 704}, 113 (2011).

\bibitem{Bazzi:2012eq}
M.~Bazzi {\em et~al.} (SIDDHARTA Collaboration),
\newblock Nucl. Phys. A {\bf 881}, 88 (2012).

\bibitem{Meissner:2004jr}
U.~G.~Meissner, U.~Raha and A.~Rusetsky,
\newblock Eur. Phys. J. C {\bf 35}, 349 (2004).

\bibitem{Ikeda:2011pi}
Y.~Ikeda, T.~Hyodo and W.~Weise,
\newblock Phys. Lett. B {\bf 706}, 63 (2011).

\bibitem{Ikeda:2012au}
Y.~Ikeda, T.~Hyodo and W.~Weise,
\newblock Nucl. Phys. A {\bf 881}, 98 (2012).

\bibitem{Yamaguchi:2016kxa}
Y.~Yamaguchi and T.~Hyodo,
\newblock Phys. Rev. C {\bf 94}, 065207 (2016).

\bibitem{Kaiser:1995eg}
N.~Kaiser, P.~B.~Siegel and W.~Weise,
\newblock Nucl. Phys. A {\bf 594}, 325 (1995).

\bibitem{Oset:1998it}
E.~Oset and A.~Ramos,
\newblock Nucl. Phys. A {\bf 635}, 99 (1998).

\bibitem{Oller:2000fj}
J.~A.~Oller and U.~G.~Meissner,
\newblock Phys. Lett. B {\bf 500}, 263 (2001).

\bibitem{Hyodo:2008xr}
T.~Hyodo, D.~Jido and A.~Hosaka,
\newblock Phys. Rev. C {\bf 78}, 025203 (2008).

\bibitem{Hyodo:2007jq} 
T.~Hyodo and W.~Weise,
Phys.\ Rev.\ C {\bf 77}, 035204 (2008).

\bibitem{Eden:1964zz}
R.~Eden and J.~Taylor,
\newblock Phys. Rev. {\bf 133}, B1575 (1964).

\bibitem{Pearce:1988rk}
B.~C.~Pearce and B.~F.~Gibson,
\newblock Phys. Rev. C {\bf 40}, 902 (1989).

\bibitem{Cieply:2016jby}
A.~Cieply, M.~Mai, U.-G.~Meissner and J.~Smejkal,
\newblock Nucl. Phys. A {\bf 954}, 17 (2016).

\bibitem{Hyodo:2014bda}
T.~Hyodo,
\newblock Phys. Rev. C {\bf 90}, 055208 (2014).

\bibitem{Kamiya:2017pcq} 
Y.~Kamiya and T.~Hyodo,
\newblock Phys.\ Rev.\ D {\bf 97}, 054019 (2018).
  
\bibitem{Hammer:2001ng}
H.~W.~Hammer,
\newblock Nucl. Phys. A \textbf{705}, 173 (2002).

\bibitem{AB10b}
S.-I.~Ando and M.~C.~Birse,
\newblock J. Phys. G \textbf{37}, 105108 (2010).

\bibitem{AYO13}
S.-I.~Ando, G.-S.~Yang, and Y.~Oh,
\newblock Phys. Rev. C \textbf{89}, 014318 (2014).

\bibitem{AO14}
S.-I.~Ando and Y.~Oh,
\newblock Phys. Rev. C \textbf{90}, 037301 (2014).

\bibitem{Hammer:2017tjm}
H.~W.~Hammer, C.~Ji, and D.~R.~Phillips,
\newblock J. Phys G {\bf 44}, 103002 (2017) 

\bibitem{Kaplan:1998tg}
D.~B.~Kaplan, M.~J.~Savage and M.~B.~Wise,
\newblock Phys. Lett. B {\bf 424}, 390 (1998).

\bibitem{Kaplan:1998we}
D.~B.~Kaplan, M.~J.~Savage and M.~B.~Wise,
\newblock Nucl. Phys. B {\bf 534}, 329 (1998).

\bibitem{vanKolck:1998bw}
U.~van~Kolck, 
\newblock Nucl. Phys. A {\bf 645}, 273-302 (1999).

\bibitem{BS01}
S.~R. Beane and M.~J. Savage, 
\newblock Nucl. Phys. A {\bf 694}, 511 (2001).

\bibitem{AH04}
S.-I.~Ando and C.~H. Hyun,
\newblock Phys. Rev. C {\bf 72}, 014008 (2005).  

\bibitem{BHV98}
P.~F.~Bedaque, H.~W.~Hammer, and U.~van~Kolck,
\newblock Nucl. Phys. A \textbf{646}, 444 (1999).

\bibitem{STM1}
G.~V.~Skornyakov and K.~A.~Ter-Martirosyan, 
\newblock Sov. Phys. JETP {\bf 4}, 648 (1957) [Zh. Eksp. Teor. Fiz. 31, 775 (1956)].

\bibitem{STM2}
G.~V.~Skornyakov and K.~A.~Ter-Martirosyan,
\newblock JETP {\bf 31}, 775 (1956).

\bibitem{Canham:2008jd}
D.~L.~Canham and H.~W.~Hammer,
\newblock Eur. Phys. J. A {\bf 37}, 367 (2008).

\bibitem{DL63}
G.~S.~Danilov and V.~I.~Lebedev,
\newblock Sov. Phys. JETP 17 (1963) 1015.

\bibitem{Chen2008}
Q.~Chen et al.,
\newblock Phys. Rev. C \textbf{77}, (2008), 054002-1.




\end{thebibliography}

\end{document}